\documentclass[preprint,journal]{vgtc}            

\onlineid{0}

\ieeedoi{10.1109/TVCG.2024.3456316}

\vgtccategory{Research}

\vgtcpapertype{Evaluation}

\title{Evaluating Force-based Haptics for Immersive Tangible Interactions with Surface Visualizations}

\author{%
  \authororcid{Hamza Afzaal}{0009-0004-8066-0523},
  and \authororcid{Usman Alim}{0000-0003-4834-2475}
}

\authorfooter{
\item 
The authors are with the Department of Computer Science, University of Calgary. E-mail: \{hamza.afzaal, ualim\}@ucalgary.ca.
}

\abstract{%

Haptic feedback provides an essential sensory stimulus crucial for interaction and analyzing three-dimensional spatio-temporal phenomena on surface visualizations. Given its ability to provide enhanced spatial perception and scene maneuverability, virtual reality (VR) catalyzes haptic interactions on surface visualizations. Various interaction modes, encompassing both mid-air and on-surface interactions---with or without the application of assisting force stimuli---have been explored using haptic force feedback devices. In this paper, we evaluate the use of on-surface and assisted on-surface haptic modes of interaction compared to a no-haptic interaction mode. A force-based haptic stylus is used for all three modalities; the on-surface mode uses collision based forces, whereas the assisted on-surface mode is accompanied by an additional snapping force. We conducted a within-subjects user study involving fundamental interaction tasks performed on surface visualizations. Keeping a consistent visual design across all three modes, our study incorporates tasks that require the localization of the highest, lowest, and random points on surfaces; and tasks that focus on brushing curves on surfaces with varying complexity and occlusion levels. Our findings show that participants took almost the same time to brush curves using all the interaction modes. They could draw smoother curves using the on-surface interaction modes compared to the no-haptic mode. However, the assisted on-surface mode provided better accuracy than the on-surface mode. The on-surface mode was slower in point localization, but the accuracy depended on the visual cues and occlusions associated with the tasks. Finally, we discuss participant feedback on using haptic force feedback as a tangible input modality and share takeaways to aid the design of haptics-based tangible interactions for surface visualizations. 

}


\teaser{
  \centering
  \includegraphics[width=\linewidth,trim={0cm 12.6cm 0cm 0 cm},clip, alt={A view of a city with buildings peeking out of the clouds.}]{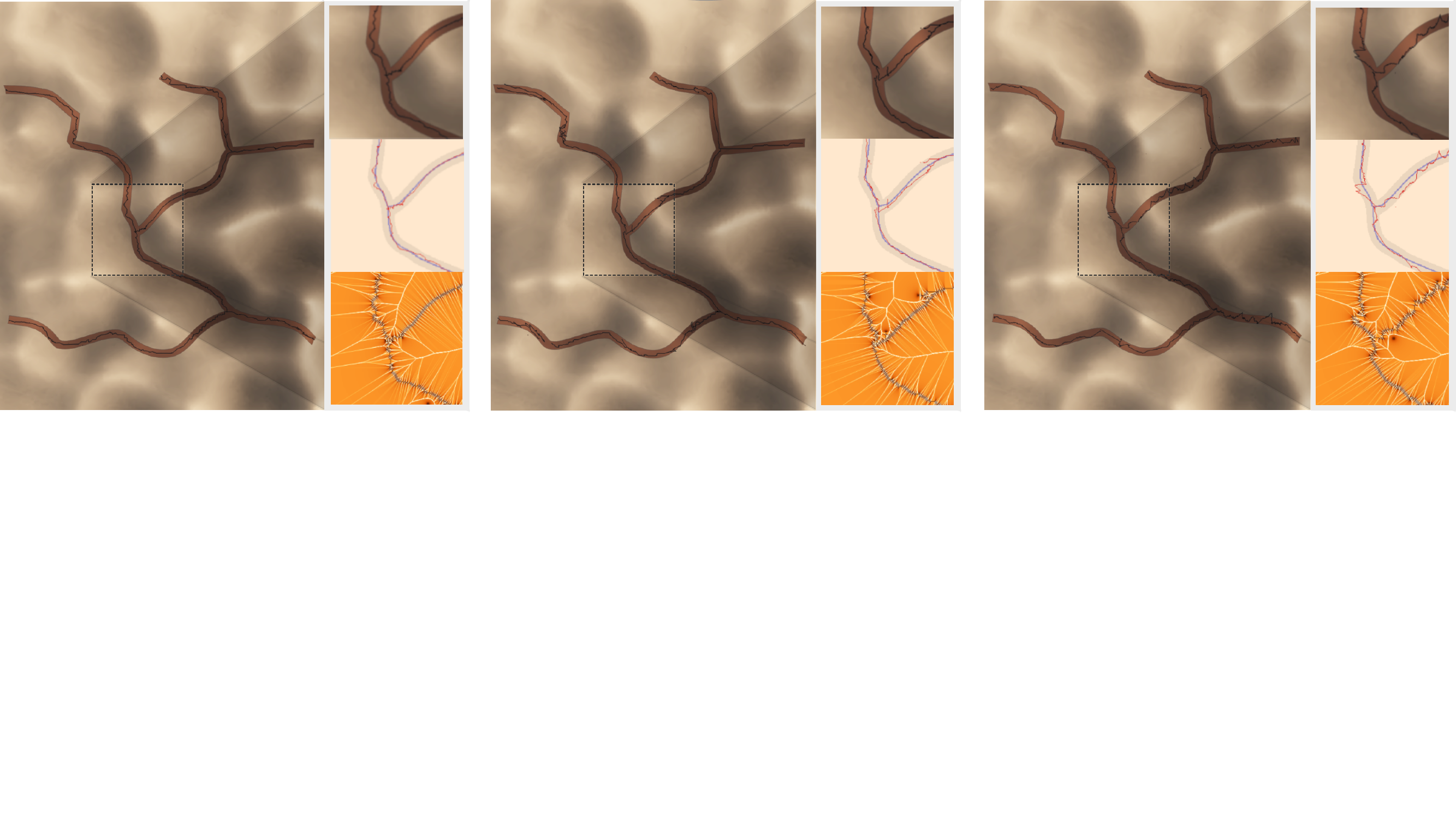}
  \caption{%
  	Participants brushed curves on a surface in VR using a haptic stylus with (left) and without (middle) snap assistance, and no-haptic feedback (right). Top to bottom: brushed curve, texture showing the brushed curve and the bands participants were asked to follow, and the Laplacian of the Euclidean distance transform of the brushed curves in texture space. Observe that participants using the haptic stylus with snapping brushed smoother and more accurate curves.%
  }
  \label{fig:teaser}
}

\graphicspath{{figs/}{figures/}{pictures/}{images/}{results/}{./}} 

\usepackage{tabu}                      
\usepackage{booktabs}                  
\usepackage{lipsum}                    
\usepackage{mwe}                       

\usepackage{array}
\usepackage{amsmath}

\usepackage[subtle]{savetrees}

\begin{document}
\definecolor{point_color}{RGB}{153, 52, 4}
\definecolor{brush_color}{RGB}{44, 127, 184}
\definecolor{revision_color}{RGB}{0, 0, 0}
\definecolor{revision_color_b}{RGB}{0, 0, 0}
\definecolor{sig_color}{RGB}{150, 150, 150}
\newcommand{\pointc}[1]{{\textcolor{point_color}{\texttt{{#1}}}}}
\newcommand{\brushc}[1]{{\textcolor{brush_color}{\texttt{{#1}}}}}
\newcommand{\sigc}[1]{{\textcolor{sig_color}{\texttt{{#1}}}}}
\newcommand{\revisionc}[1]{{#1}}
\newcommand{\revisionb}[1]{{#1}}


\firstsection{Introduction}
\maketitle

Surface visualizations are pivotal for analyzing, exploring, and interpreting 3D shapes and properties from medical imaging, computer-aided design, and geospatial datasets\cite{edmunds_surface-based_2012, westermann_real-time_1999, mills_geospatial_2008, tietjen_combining_2005, lawonn_survey_2018}. Interpreting such visualizations requires an understanding of the surface's depth, distance, and orientation \cite{gibson_perception_1950, ware_view_nodate}. Virtual reality (VR) has improved the comprehension of these visualizations \cite{fonnet_survey_2021} by providing visual cues such as field of regard, stereoscopy, and head tracking for better viewing angles and depth perception \cite{laha_effects_2014, laha_effects_2012}. Still, perception and interaction with surface visualizations in VR depend on some intrinsic properties like shading, lighting and texturing to better understand the surface's depth and shape \cite{ware_view_nodate, lawonn_improving_2017}. Moreover, surface visualizations inherit occlusion problems due to region overlaps; hence the view has to be adjusted continuously to gain a better perspective of the entire visualization \cite{meuschke_evalviz_2019, elmqvist_taxonomy_2008, preim_survey_2016}. This continuous adjustment of the view affects the memorability aspect of the visualization, increasing the cognitive overhead needed to switch back-and-forth between regions of interest on a surface \cite{alabi_comparative_2012}.

Researchers have explored different interaction modalities in combination with VR to provide users with additional sensory stimuli for navigating and understanding surface visualizations \cite{besancon_state_2021}. Haptification of 3D spatial data using a force-based input device is widely used in the scientific community to provide a touch sensation from the visualized data \cite{lin_haptic_2008}. These tangible input devices provide an additional channel of information for perceiving visual cues like depth and shape of the visualizations by activating the lateral optical visuo-tactile area of the brain, hence complementing the interpretation of the visualizations \cite{reiner_seeing_2008}. Moreover, the physical perception of the surfaces with force-based haptics having six degrees of freedom (DoF) facilitates the interaction with occluded and cluttered regions \cite{besancon_state_2021}. Some have explored the use of assistive snapping forces along with the haptic devices for planar \cite{mohanty_investigating_2020} or complex surface interactions \cite{laehyun_kim_haptic_2004} to provide additional stimuli for selection and navigation.


Considering the significance of surface visualizations in the scientific community and the diverse range of force-based haptic modes employed for interactions, ranging from mid-air drawing to on-surface sketching with or without force assistance, this paper aims to explore how the enhanced depth and shape perception offered by integrating VR with force-based haptics ---\emph{with or without assistance}--- impacts user performance in interactive surface visualization tasks. To the best of our knowledge, this is the first paper that presents a comparative study contrasting three modes for surface interaction tasks: a visual-only mode with no haptics, an on-surface force-based haptics mode that uses collision-based forces, and an on-surface force-based haptics mode that is combined with an additional assistive force that snaps the haptics device to the surface. To assist force-based haptics, we also introduce a novel force profile that allows for smoother snapping and ease of maneuverability on the surface. 

To accomplish our goal, we designed and conducted a quantitative user study that employs common visual interaction tasks such as point localization and brushing curves on surface visualizations~\cite{li_visibility_2017,zhou_haptics-assisted_2008}. 
Our findings elucidate scenarios where the force-based haptics, with or without force assistance, assists interaction with surface visualizations (see \cref{fig:teaser}), and scenarios where visual-only cues are sufficient to accomplish the tasks. 
\revisionb{We also present design guidelines for future researchers on employing force-based haptics in surface-based visualizations.}

Our contributions can be summarized as follows: 
\begin{itemize}[topsep=0pt,itemsep=0pt,partopsep=0pt, parsep=0pt]
    \item A directional \textbf{snapping force} based on the distance transform of a surface and a distance-based force profile facilitating tangential movements on the surface.
    
    \item \revisionb{Empirical results from a \textbf{user study} comparing user performance on six interaction tasks with surface visualizations. Twenty-four participants localized the highest, lowest, and random points, and brushed curves on surfaces of varying complexity. We interpret these results to provide design guidelines for developing interactive visualization systems that incorporate force-based haptics.}
    \item A set of \textbf{takeaways} and lessons learned from the user study based on feedback received from the participants.
\end{itemize}

\section{Related Work}
\label{sec:related_work}


In this section, we review previously used techniques for surface visualizations and discuss their broad-spectrum use cases and applications. We also explore different techniques for interaction that rely on visual cues from the surface. Furthermore, we review the use of tangible and hybrid input modalities for interacting with surface visualizations to establish a baseline for our surface visualization design strategy. 

\subsection{Importance of Surface Visualization}
Researchers have utilized various techniques to visualize 3D data, aiming to decipher its complex characteristics. Visualizations of volumetric data that illustrate complex surfaces through scalar values assigned to voxels, often employ the marching cubes algorithm to extract isosurfaces by identifying specific scalar values and conducting triangulation on the voxels~\cite{lorensen_marching_1987}. Sunderland \textit{et al}.~\cite{webster_reconstruction_2015} presented a technique for building surface-based visualizations from 2D contours extracted from slices in volumetric data. The visualization of parametric surfaces is pivotal in visualizing mathematical models and is an essential component of Computer Aided Design~\cite{kinnear_procedural_2010}. Applications of surface visualizations are also found in geospatial data such as digital elevation models (DEM) \cite{zhou_digital_2017, moore_digital_1991}.

\revisionb{Surface visualizations extend beyond visualizing scalar data or mathematical models, serving to present additional information through various encodings. Multivariate visualizations encode data on surfaces using colors and textures, benefiting applications such as fluid simulations\cite{englund_touching_2018}, and decal-based techniques proposed by Rocha \textit{et al} ~\cite{rocha_decal-maps_2017,rocha_decal-lenses_2019}.} 
Tietjen \textit{et al}.~\cite{tietjen_combining_2005} used silhouettes, volumes, and surfaces to depict components in a volumetric dataset with greater detail. Ropinski \textit{et al}.~\cite{ropinski_visually_2006} proposed a depth-based color encoding technique that heightens depth perception in surfaces such as blood vessels. Alabi \textit{et al}.~\cite{alabi_comparative_2012} proposed a method to superimpose surfaces to find similarities and differences between ensembles. Lawonn \textit{et al}.~\cite{lawonn_survey_2018} review works that use illustrative techniques to improve the perception of surface-based visualizations. 

\revisionc{The review of surface visualizations lays a foundation for selecting appropriate techniques for designing the snapping force and user study. 
Their understanding and applications aids us in creating interaction tasks that effectively incorporate multivariate cues and visual representations.}

\subsection{Interaction Techniques for Surface Visualization}
Research on interaction with surface visualizations spans from traditional 2D displays to immersive VR environments. Ohnishi \textit{et al.} \cite{ohnishi_virtual_2012} introduced virtual interaction surfaces in 2D environments for 3D surface manipulation, including pointing, placement, and texture mapping. Wang \textit{et al.} \cite{wang_understanding_2022} evaluated 2D and 3D input/output devices for interacting with 3D visualizations, highlighting the advantages of each approach. Laha \textit{et al.} \cite{laha_effects_2014} assessed VR's fidelity for analyzing isosurfaces through tasks like pattern recognition and spatial judgment. Meuschke \textit{et al.} \cite{meuschke_evalviz_2019} conducted a study on perceptual identification of shapes and depths in surface visualizations. Song \textit{et al.} \cite{song_wysiwyf_2011} utilized tangible input devices, such as an iPad, for selecting, slicing, and annotating volumetric data. Usher \textit{et al.} \cite{usher_virtual_2018} employed VR handheld controllers for neuron tracing on isosurface reconstructions of primate visual cortex microscopy scans.

Another line of research has explored tangible inputs for performing complex manipulations on 3D data~\cite{besancon_state_2021}. In the realm of mid-air interactions, applications have extended to sketching and modelling, often involving curvature-based movements. Choi \textit{et al.}~\cite{choi_force_2005} proposed a constancy hypothesis for virtual surface interaction highlighting that a continuous haptic stimulus induces fatigue; thereby favouring mid-air interaction methods where haptic stimuli are sparingly used. Prouzeau \textit{et al.}~\cite{prouzeau_scaptics_2019} explored the use of VR controllers for exploring 3D scatter plots using highlight planes to filter data. Machucha \textit{et al.}~\cite{machuca_multiplanes_2018} used VR controllers to draw strokes on arbitrary planes placed in space. \revisionb{Additionally, pen and tablet-based approaches have been used for more direct interactions with surfaces~\cite{arora_symbiosissketch_2018,kim_agile_2018,bae_ilovesketch_2008}, facilitating intuitive sketching and modelling processes with vibrotactile feedback.} Moreover, some researchers have adopted probes equipped with vibrotactile, thermal, or pneumatic actuators, to support immersive mid-air drawing tasks~\cite{jackson_lift-off_2016,hoffmann_thermalpen_2023,grossman_creating_2002, strohmeier_generating_2017}. 

However, force-enabled haptic interactions find applications in tasks involving precision and finer manipulations~\cite{wang_haptic_2014, lin_haptic_2008, avila_haptic_1996, ikits_constraint-based_2003, paneels_review_2010}.  Basdogan \textit{et al.}~\cite{basdogan_haptic_2001} discuss the applications of haptic interaction in immersive environments. Englund \textit{et al.}~\cite{englund_touching_2018} used visuo-haptic visualizations to explore fluid flow by interacting and relaying sensory information through force-based haptics from isosurfaces. Corenthy \textit{et al.}~\cite{corenthy_volume_2015} explored using force-based haptics to interpret topology-consistent isosurfaces. Zhou \textit{et al.}~\cite{zhou_haptics-assisted_2008} used 6-DoF from force-based haptic devices to draw lassos to select annotation regions on surfaces generated from diffusion tensor imaging. Some researchers utilized assistive forces such as snap and magnets to assist with surface interactions. Jackson \textit{et al.}~\cite{jackson_force_2012} used a snapping force with a haptic device to analyze flow visualizations. Laehyun \textit{et al.}~\cite{laehyun_kim_haptic_2004} used snapping forces on surfaces generated from implicit data for sketching. Komerska \textit{et al.}~\cite{komerska_haptic_2004} used snap and pop through features for surface-based interactions. Keefe \textit{et al.}~\cite{keefe_drawing_2007} proposed mid-air curve drawing using one-handed and two-handed tape drawing methods. Mohanty \textit{et al.}~\cite{mohanty_investigating_2020} presented a double-suction approach for interacting with arbitrary placed mid-air planes for curve sketching. Top \textit{et al.}~\cite{top_spotlight_2011} presented a method for active assistance along with haptic force feedback to steer the user through annotated regions on 3D surfaces. Zhang \textit{et al.}~\cite{zhang_haptic_2023} presented an approach for haptic interactions with neural radiance field data. 


While the literature features a variety of haptic interaction modes and their applications, 
comprehensive comparisons of these modes across different surface visualization and interaction tasks are notably lacking.
Additionally, despite the existence of numerous surface-based haptic interaction approaches, the detailed implementation of snap-to-surface assistive forces for complex explicit surfaces are rarely shared. Existing studies frequently employ assistive force profiles based on Hookean forces, which may lead to overly aggressive detachment or inadequate snapping forces, compromising the user experience. This paper intends to provide a detailed implementation of a smoother force profile for snapping assistance on arbitrary surfaces. 
We also investigate the effects of employing domain-agnostic haptics ---\textit{with or without force assistance}--- relative to a no-haptic mode across standard interaction tasks.

\section{Forcing Methodology}

\revisionb{This section outlines the methodology for computing haptic and haptic snap forces within the context of physically-based dynamics. 
The OpenHaptics\textsuperscript{\sffamily\textregistered} SDK~\cite{itkowitz_openhaptics_2005} was employed for interfacing with the haptic device, enabling implementation of our method using
position-based dynamics to enhance the accuracy of haptic feedback. Parameter selection is also discussed, providing insight into the underlying rationale.}


\subsection{Haptic Rendering}

We adopted a proxy-based haptic rendering technique for interacting with surfaces \cite{zilles_constraint-based_1995}. In this method, the position of a proxy $\mathbf{p}$, also known as a god-object, is closely tracked along with the position of the haptic stylus $\mathbf{s}$. During interaction with a surface $M$, the proxy stays on the surface, while the actual position of the haptic stylus $\mathbf{s}$ is allowed to penetrate the surface. A force $\mathbf{F}_{k}$ is subsequently applied to the actual stylus via a virtual spring-damper mechanism that links $\mathbf{p}$ and $\mathbf{s}$ (see~\cref{fig:snap_to_surface}). This force is given by
\vspace{-6px}
\begin{equation}
\mathbf{F}_k := -\kappa (\mathbf{s-p}) - \tau\,\dot{\mathbf{s}},  
\end{equation}
\vspace{-15px}

\noindent where $\kappa$ represents the spring constant, $\dot{\mathbf{s}}$ is the velocity of the stylus and $\tau$ is a damping coefficient. 

During a pilot study (Sec.~\ref{sec:pilot_study}), we adjusted the values for $\kappa$ and $\tau$. The employed SDK also implements surface friction and air viscosity following the extension of the proxy method by Ruspini \textit{et al.}~\cite{ruspini_haptic_1997}. These values were also fine-tuned during the pilot study to provide an optimal feel while navigating and interacting with surfaces.
\revisionc{The forces computed through the proxy constraint method have been proven effective for point-to-rigid-body interactions, demonstrating realistic haptic feedback \cite{lorenz_perceived_2023}. Building on this approach, our methodology simulates interactions between rigid surfaces and a rounded steel rod, aligning well with interactions involving hard, non-deformable surfaces while mirroring real-life tactile experiences. We gather user feedback to further enhance realism by fine-tuning surface hardness and haptic feedback parameters.}

\subsection{Haptic Snap Rendering}
We detail a technique for applying an assistive snapping force to the haptic rendering process. This force activates when the haptic stylus enters a predefined `snap distance' from the surface, directing the stylus towards the nearest point on the surface $M$. The method for calculating the snap distance and the corresponding force profile, which pulls the stylus towards $M$, are outlined below.

\begin{figure}[t]
  \centering 
  \includegraphics[width=\columnwidth, alt={A visual showing snap-to-surface and force profiles}]{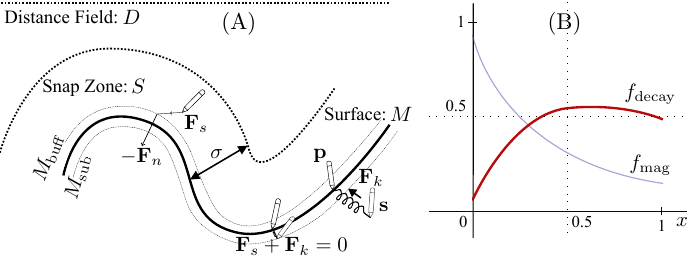}
  \caption{%
  	A visual depiction of forces applied on the haptic stylus inside the snap zone (A), and different force profiles (B). %
  }
  \label{fig:snap_to_surface}
\end{figure}

\paragraph{\textbf{Snap Distance}}
As the stylus proxy $\mathbf{p}$ approaches the surface $M$, a snapping force $\mathbf{F}_{s}$ is activated upon crossing a predefined distance threshold $\sigma$ which delineates a snap zone $S$ above $M$ as shown in~\cref{fig:snap_to_surface}. Several approaches have been discussed in the literature that utilize such forces for surface interaction, primarily through a mass-spring model, providing mid-air haptic interactions on 2D planes~\cite{elsayed_vrsketchpen_2020, mohanty_investigating_2020, keefe_drawing_2007, machuca_multiplanes_2018, grossman_creating_2002}. In our case, since we are dealing with an arbitrary surface $M$, we employed the distance field $D$ of surface $M$. The scalar distance field $D$ allows us to compute the direction of the force that is to be applied to the haptic stylus. Further coupled with a distance-based magnitude profile, $D$ gives us the desired force needed for the snapping action. 

To approximate $D$ for a surface $M$ given as a mesh, we initially calculated the bounding box of $M$ and \revisionc{isotropically scaled it by a factor of $1.4$ to serve as the enclosing volume. The scale value was determined experimentally to work with all the surfaces employed, ensuring enough room to accommodate the snap zone $S$ above a given surface $M$. 
} This enclosing volume was then subdivided into a voxel grid. Each voxel in this grid stores the distance to the nearest point on $M$, thereby approximating $D$. 
The direction to the nearest point on the surface can then be obtained via $-\nabla D$, the negative gradient of $D$.

To trigger the snap mechanism, we used trilinear interpolation on the voxel grid corresponding to $D$ to obtain the distance between the stylus proxy $\mathbf{p}$ and $M$. Upon crossing a predetermined threshold $\sigma$ (i.e., when the proxy enters snap zone $S$), we retrieve both the distance and the direction to the nearest point on $M$.
Within the snap zone $S$, we applied a snapping force $\mathbf{F}_{s}$ to the proxy, pulling it towards the nearest point on the surface $M$. 
 When the proxy is near the surface within a buffer zone $M_{\text{buff}}$ (\cref{fig:snap_to_surface}), the direction of the force is switched to $-\mathbf{F}_{n}$, where $\mathbf{F}_{n}$ denotes the normal to the surface $M$ at the nearest point. This reduces the unintended jittery motion of the stylus due to the error in approximating $D$ near $M$.
The strength of the snapping force within the snap zone $S$ is described in the following section.

\paragraph{\textbf{Snap Force Profiling}}
Upon entering the snap zone $S$, the stylus is attracted to the surface with a variable force profile. Various methods, including those based on a Hookean spring force model (also known as magnetic force), have been documented for facilitating snapping, offering a consistent stimulus to users for implicit and explicit interaction tasks~\cite{laehyun_kim_haptic_2004, jackson_force_2012, duriez_realistic_2006, faeth_combining_2008}. In these works, the force exerted on the stylus is directly proportional to its distance from the surface. Using such a force would lead to an abrupt increase once the stylus enters the snap zone. This can be mitigated by reducing 
the spring constant, but that would also weaken the force within the snap zone.   
Bearing these considerations in mind, we employed a non-linear decaying function (see~\cref{fig:snap_to_surface}) given by
\vspace{-6px}
\begin{equation}
    f_{\text{decay}}(x) = A\,g(x)\,e^{-Bg(x)}, \quad \text{where} \quad
   g(x) = 0.78x + 0.02.
\end{equation}
\vspace{-15px}

\noindent Here, $A$ is the amplitude and $B$ is the decay rate. The magnitude of the snapping force $\mathbf{F}_s$ is then given by $f_{\text{decay}}(r/\sigma)$ where $r$ is the distance of the proxy $\mathbf{p}$ from the surface $M$.

Upon entering the snap zone $S$, the magnitude of $\mathbf{F}_{s}$ gently increases, followed by a gradual decrease. This ensures that, when the stylus makes contact with the surface $M$, the pulling force is minimal, facilitating smooth movement of the stylus over the surface. Similarly, as the user lifts the stylus away from $M$, the force profile mimics a mass-spring mechanism, offering a force increment to prevent unintended disengagement from the surface. Ultimately, as the stylus approaches the no-snap zone from within the snap zone, the force diminishes gradually, ensuring a smoother transition out of the snap influence.

When the proxy rests on the surface $M$, there is a small snapping force $\mathbf{F}_{s}$ that pulls the actual stylus below the surface $M$. A counteracting force from the mass-spring force $\mathbf{F}_{k}$ pushes the stylus back, thereby creating an equilibrium state underneath the surface. The locus of points where this equilibrium is achieved is denoted as $M_{\text{sub}}$ in~\cref{fig:snap_to_surface}. This ensures that the user finds the surface sticky when the stylus snaps to the surface but can easily move the stylus in tangential directions. 

\revisionc{The snapping force simulates a weak magnetic attachment and a spring effect when lifting the stylus. It is important to note that the force does not resemble any physically-based force. Similar to illustrative visualization which is based on non-photorealistic rendering techniques \cite{lawonn_survey_2018}, we propose that a departure from a physically-based force model may enhance immersive interactions on surface visualizations.}

\subsection{Pilot Study}
\label{sec:pilot_study}

\textbf{Objective:}
We designed our pilot study to fine-tune our force profile function. In addition, we compared our decaying non-linear force profile with one based on the inverse square law. We opted for this force profile as a control group as it exerts an increasing force---characterized by ${1/(2x+1)^2}$ and denoted as $f_{\text{mag}}$ in~\cref{fig:snap_to_surface}--on the stylus within the snap region, fostering a stronger connection between the virtual stylus and the surface. 
\revisionb{Our evaluation focused on whether a gradual force increase upon entry and a stronger force upon contact provides a better experience than a slightly stronger force upon entry but a weaker force upon contact. The goal was to find the force profile with fine-tuned parameters that feels comfortable and helps in precise and stable navigation across the features of a surface visualization.}


\textbf{Participants:}
\revisionc{We recruited three participants (2 male, 1 female) from research labs focusing on scientific visualization. Two participants had experience using a haptic pen device in their research, while the third was a novice. All participants were familiar with using a VR headset and were comfortable employing it for the pilot study.}

\textbf{Methods:}
For our study, we selected a surface visualization with significant topographic variations to examine the tactile feedback during snap-to-surface haptic interactions across both protruded and recessed areas. We manually adjusted the force profiles and experimented with various parameters \revisionc{(amplitude \(A\) and decay rate \(B\) were varied on a linear scale from 1 to 5 in 0.5 increments)}, instructing the participants to perform three specific tasks: i) move the virtual stylus close enough to trigger the snap-to-surface feature, ii) lift the stylus away from the surface to a distance beyond the snap zone, and finally, iii) glide the stylus in a circular motion over areas of topographic relief. Throughout the experiment, we collected their qualitative feedback on each interaction mode and its associated parameters.

\textbf{Feedback \& Adjustments:}
Participants' feedback highlighted that, although the inverse-square force profile facilitated a smoother transition between snap and no-snap regions, the increased force exerted by the haptic device during surface contact made navigating over irregular topographies, such as bumps and crevices, more challenging. This difficulty was attributed to the stylus's tendency to slide aggressively on inclined surfaces and the added effort needed to lift the stylus away from the surface. 
In contrast, participants found that the decaying non-linear force profile offered a gentle pull as the stylus entered the snap zone, enabling smoother movement close to the surface. 
\revisionc{Participants could toggle between parameters and request adjustment of values in real-time to achieve optimal feel. For decay rate, a lower value led to an abrupt snap, while a higher value decreased initial attraction force. Participants favored an average decay rate of 2 for a smooth, yet strong snap. Similarly, higher amplitude increased snapping force near the surface, while lower values had the opposite effect. An average amplitude of 3 was preferred for a strong yet controllable snap, leading to easier tangential movement on surfaces.}

\revisionc{We used user feedback to refine the haptic rendering parameters. Users preferred a stiffer surface, leading us to increase the spring stiffness and damping coefficients, which provided a solid foundation for haptic feedback. They also favored low surface friction for better traction, noting that it made the stylus easier to glide when assisted, while higher friction was fatiguing. Users wanted to eliminate air friction across all modes, as it caused continuous haptic stimuli that were tiring, consistent with findings reported by Strohmeier \textit{et al.}~\cite{strohmeier_generating_2017}.}


\section{Design Choices for Evaluation}
\label{sec:design_choices}
We now elaborate on the choices made for our surface visualizations and selection of other interaction techniques for comparison against snap-to-surface haptic interaction. 

\subsection{Haptic Design}

\revisionc{Our haptic design employs a GeoMagic\textsuperscript{\sffamily\textregistered} Touch\textsuperscript{\sffamily\texttrademark} device, featuring a $160 mm \times 120 mm \times 70 mm$ workspace, 3 DOF force feedback, and 6 DOF position sensing with $\sim0.055mm$ resolution \cite{geomagic}, enhancing interactions with additional tactile stimuli. This device offers adequate workspace and high resolution for surface visualizations at a lower cost, making it the preferred choice for our study.}

\revisionc{We evaluate our snap-to-surface haptic interaction technique against two other modes \emph{using the same haptic device}: haptic-only and no-haptic interactions. The haptic-only mode keeps all baseline components of snap-to-surface, except the snapping feature, requiring the stylus to touch the surface to register an action. In contrast, the no-haptic mode removes haptic feedback, relying only on visual cues from a pointer cast onto the surface. These modes depict on-surface and mid-air interaction techniques commonly used in surface visualization tasks and were separated to measure the effects of each in various interaction tasks.}

\subsection{Visual Design}

In our study, we use various surface types, designs, and rendering methods to compare three interaction styles. These choices provide a standard baseline for our investigation, ensuring fair comparisons. We create and refine surface models, set up navigation controls, and apply shading techniques systematically to maintain consistency across modalities. This approach enables us to accurately evaluate each interaction technique's effectiveness.

\paragraph{\textbf{Surface Models}}
\label{sec:surface_models}
Common visualization surfaces fall into two main types: explicit and implicit. Explicit surfaces directly outline points in three dimensions, often via a mesh representation. Implicit surfaces, on the other hand, are defined by a level-set of a 3D scalar function. We chose explicit surfaces for our study because they're commonly used in tasks involving physical sensations. They also make it easy to map function values onto surfaces for interacting with multivariate data. We obtained surface models for our study through three methods:

\begin{enumerate}
[topsep=0pt,itemsep=0pt,partopsep=0pt, parsep=0pt]
    \item \textit{Procedural surfaces }are created by merging procedural textures such as Perlin noise \cite{perlin}, 2D Gaussian and sinusoidal functions, as height maps to produce surfaces. We used randomized parameters and combined the textures using weighted blending to generate surfaces exhibiting a variety of features.
    \item \textit{Hand sculpted surfaces} are created using the sculpt tool for shaping a mesh in Blender~\cite{blender}.
    The sculpt tool provided precise control over the generation of specific features on the surface. We utilized sculpting to add surface features such as depressions.
    \item \textit{Object models} that were sourced from the surface annotation benchmark dataset of Chen \textit{et al.} \cite{chen_benchmark_2009}.
\end{enumerate}

\revisionc{Inappropriate surface meshing can cause jagged artifacts during brushing, often due to highly acute angles between adjacent triangles. These acute angles lead to micro-vibrations of the haptic stylus, particularly in regions with high curvature due to normal interpolation, affecting haptic feedback. To improve mesh quality and ensure smoother stylus gliding, we upsample the height fields of 100x100 by a factor of 2 for procedural surfaces and subdivision by factor 4 for others, followed by Delaunay triangulation as described by Cazals \textit{et al.}~\cite{cazals_delaunay_2006}. We then compute smooth surface normals using Newell's method. This approach ensures higher triangle density with less acute angles in curved areas, resulting in fewer artifacts and enhanced shading effects.}

To explore interactions amid occlusion and clutter, diverse surface visualizations are crucial, as noted in \cref{sec:related_work}. Our surface models draw from existing work, emphasizing isosurfaces and parametric surfaces. Using procedural techniques allows us to control surface variability, including occlusion and clutter levels, and add depth enhancement. We investigated open/closed surface models and varying topologies to examine a broad range of shapes.


\paragraph{\textbf{Navigation Controls}}
Effective scene navigation is crucial in surface visualizations. We opted for VR due to its superior three-dimensional perception and navigation capabilities. VR's precise head tracking and wide field of view require appropriate controls for navigating occluded or cluttered regions. Our approach aligns with strategies in existing literature, like Nam \textit{et al.}'s method~\cite{nam_worlds--wedges_2019}, which uses handheld controller joysticks to translate and rotate surface visualizations. This intuitive technique facilitates immersive exploration of complex spatial environments. \revisionc{VR enhances visual and hand coupling for maneuvering a haptic stylus in 3D space, while providing a realistic sense of touch and improved maneuverability of the surface in front of the user \cite{bowman_virtual_2007}.}

We set up navigation controls on the VR controller for the non-dominant hand to avoid interfering with the haptic stylus in the dominant hand. The primary button handles point localization, the trigger button is for brushing curves, and the joystick button submits tasks. The joystick manages surface visualization transformations, enabling continuous motion in 3D space when held in a particular direction. Rotation occurs in 30° increments on the y-axis to prevent user dizziness, a concern reported by existing research~\cite{cazals_delaunay_2006}. These choices ensure consistency in approaching surface regions across interaction modalities for a fair evaluation.

\paragraph{\textbf{Rendering}}
Effective shading cues are vital for surface visualization, aiding users in perceiving depth and shape during interaction \cite{ware_view_nodate}. We applied the following techniques consistently across all visualizations:
\textit{Phong shading} \cite{claussen_real_1992}, enhancing lighting cues for shape and depth perception, and \textit{screen-space ambient occlusion} (SSAO) \cite{mittring_finding_2007}, highlighting occluded regions for clearer surface perception.
In \cref{fig:inter_tasks}, a sample rendering of a procedurally generated surface visualization illustrates these techniques applied sequentially after applying a base color to the surface. Additionally, to help users quickly locate the virtual stylus and precisely place it on the surface, we added a laser pointer overlay aligned with the haptic stylus direction, as depicted in \cref{fig:inter_tasks}.

\subsection{Interaction Design}
\label{sec:interaction_design}
\revisionb{Choosing the right interactions for surface visualization is crucial for evaluating the snap-to-surface haptic interaction against other modes.}
\revisionb{The goal is to identify interactions that highlight strengths and shortcomings of snap-to-surface mode. Besançon \textit{et al.}~\cite{besancon_state_2021} have provided a taxonomy of interactions for 3D spatial data, stressing the importance of tangible inputs in 3D data selection and annotation tasks. These include basic interactions with surface models using points, curves, and regions to highlight features. We focus on two generic surface interactions: \emph{point localization} and \emph{curve brushing}.}

\begin{figure*}[tbh]
  \centering
  \begin{subfigure}[b]{0.32\linewidth}
  	\centering
  	\includegraphics[width=\textwidth,trim={16cm 12cm 22cm 9.5cm},clip]{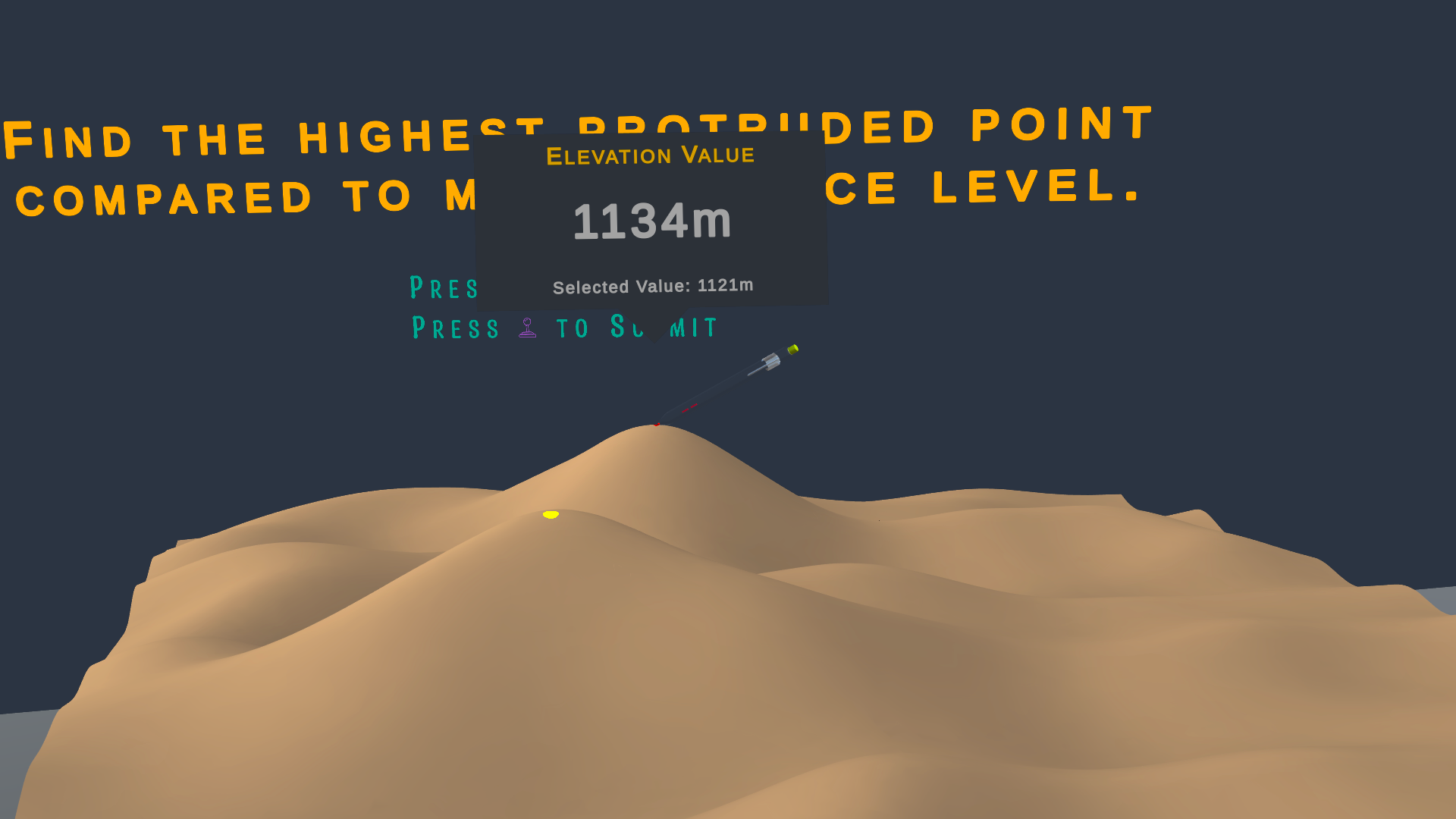}
  	\caption{Protrusion Point Localization}
  	\label{fig:protrusion}
  \end{subfigure}%
  \hfill%
  \begin{subfigure}[b]{0.32\linewidth}
  	\centering
  	\includegraphics[width=\textwidth,trim={18cm 7.5cm 20cm 14cm},clip]{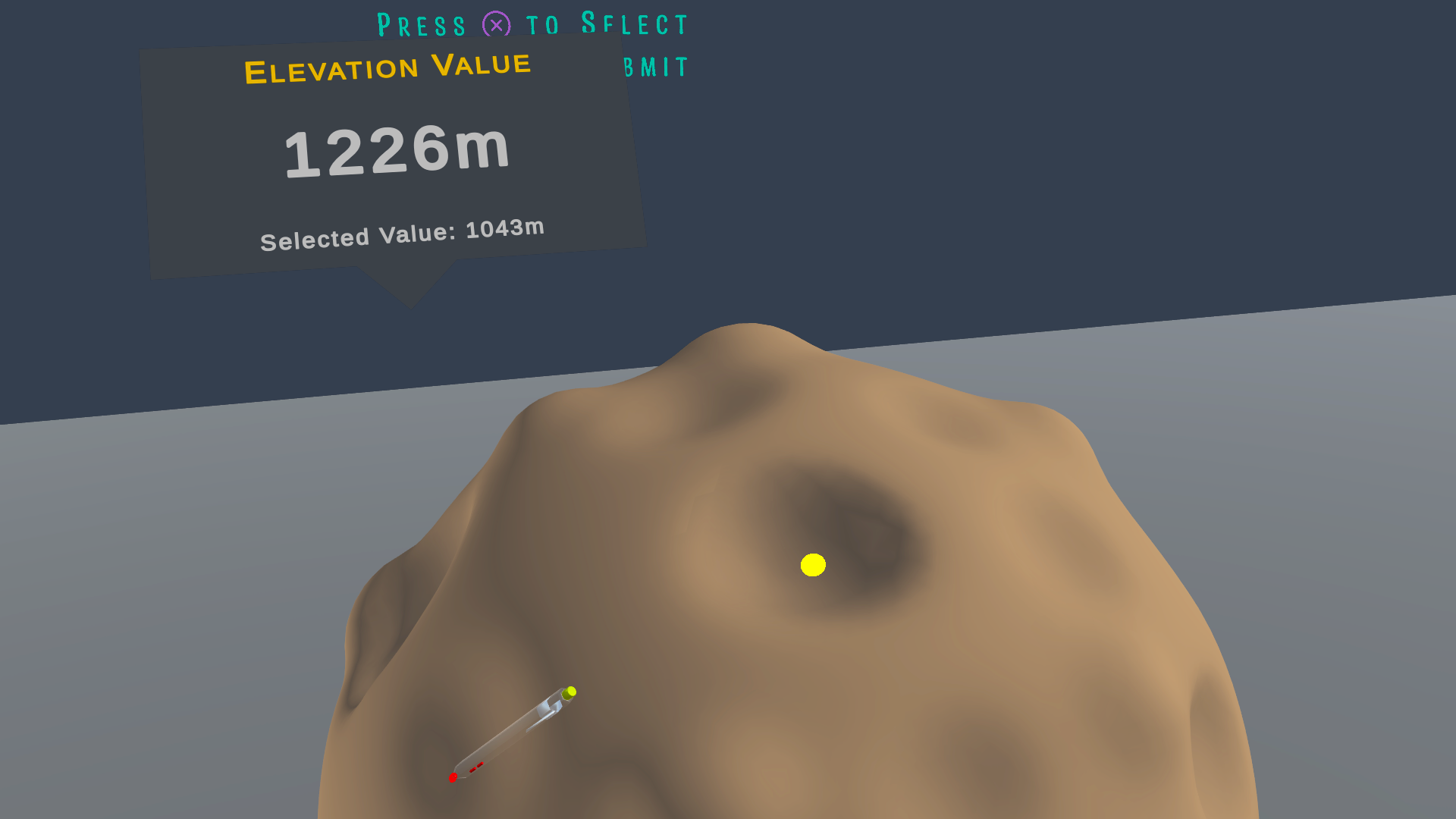}
  	\caption{Depression Point Localization}
  	\label{fig:depression}
  \end{subfigure}%
  \hfill
  \begin{subfigure}[b]{0.32\linewidth}
  	\centering
  	\includegraphics[width=\textwidth,trim={20cm 5.5cm 18cm 16cm},clip]{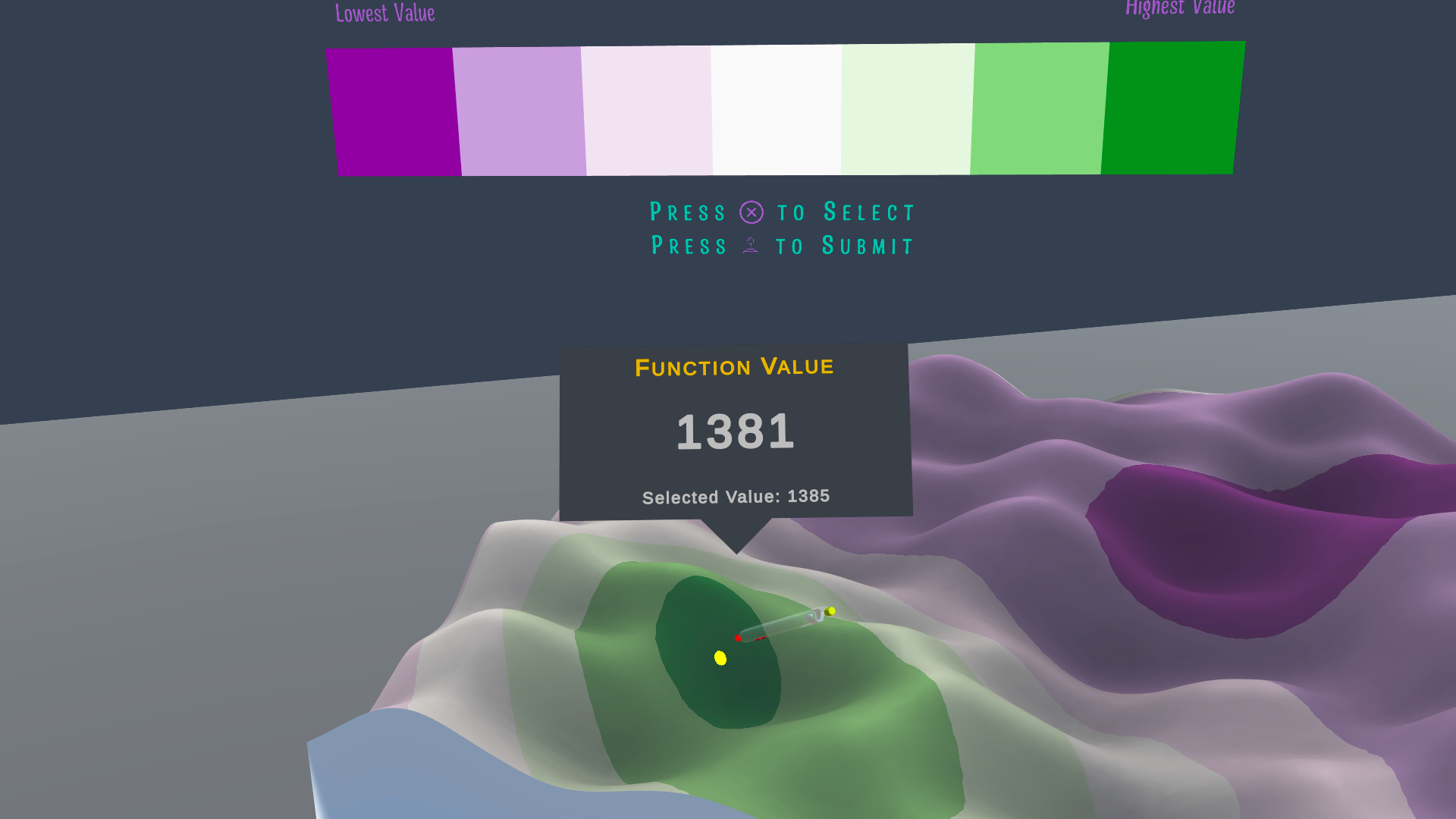}
  	\caption{Random Value Surface Point Localization}
  	\label{fig:randval}
  \end{subfigure}%
  \\
  \begin{subfigure}[b]{0.32\linewidth}
  	\centering
  	\includegraphics[width=\textwidth,trim={28cm 5cm 10cm 16cm},clip]{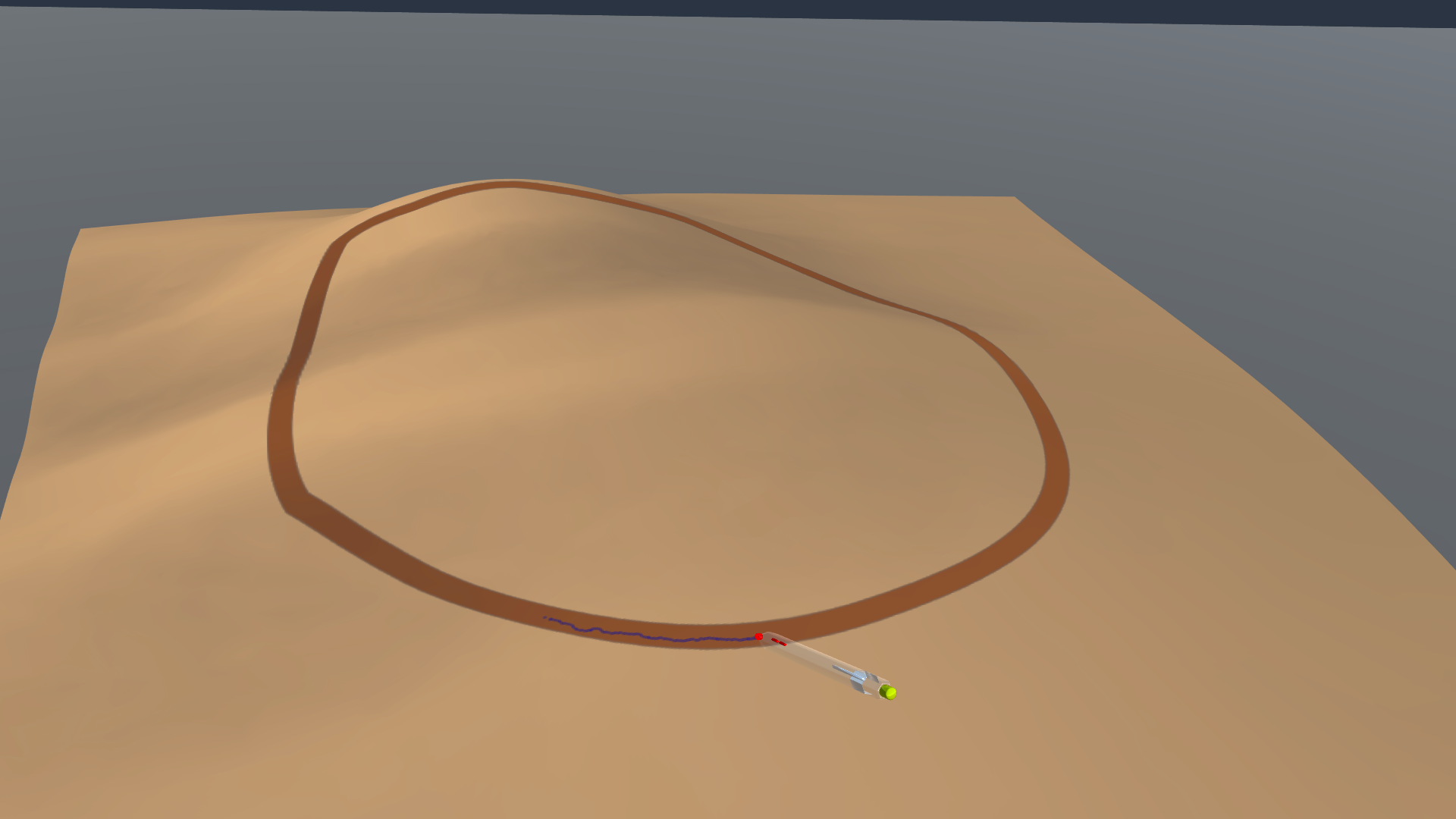}
  	\caption{Visual Curve Brushing}
  	\label{fig:contour}
  \end{subfigure}%
  \hfill%
  \begin{subfigure}[b]{0.32\linewidth}
  	\centering
  	\includegraphics[width=\textwidth,trim={28cm 5.cm 10cm 16cm},clip]{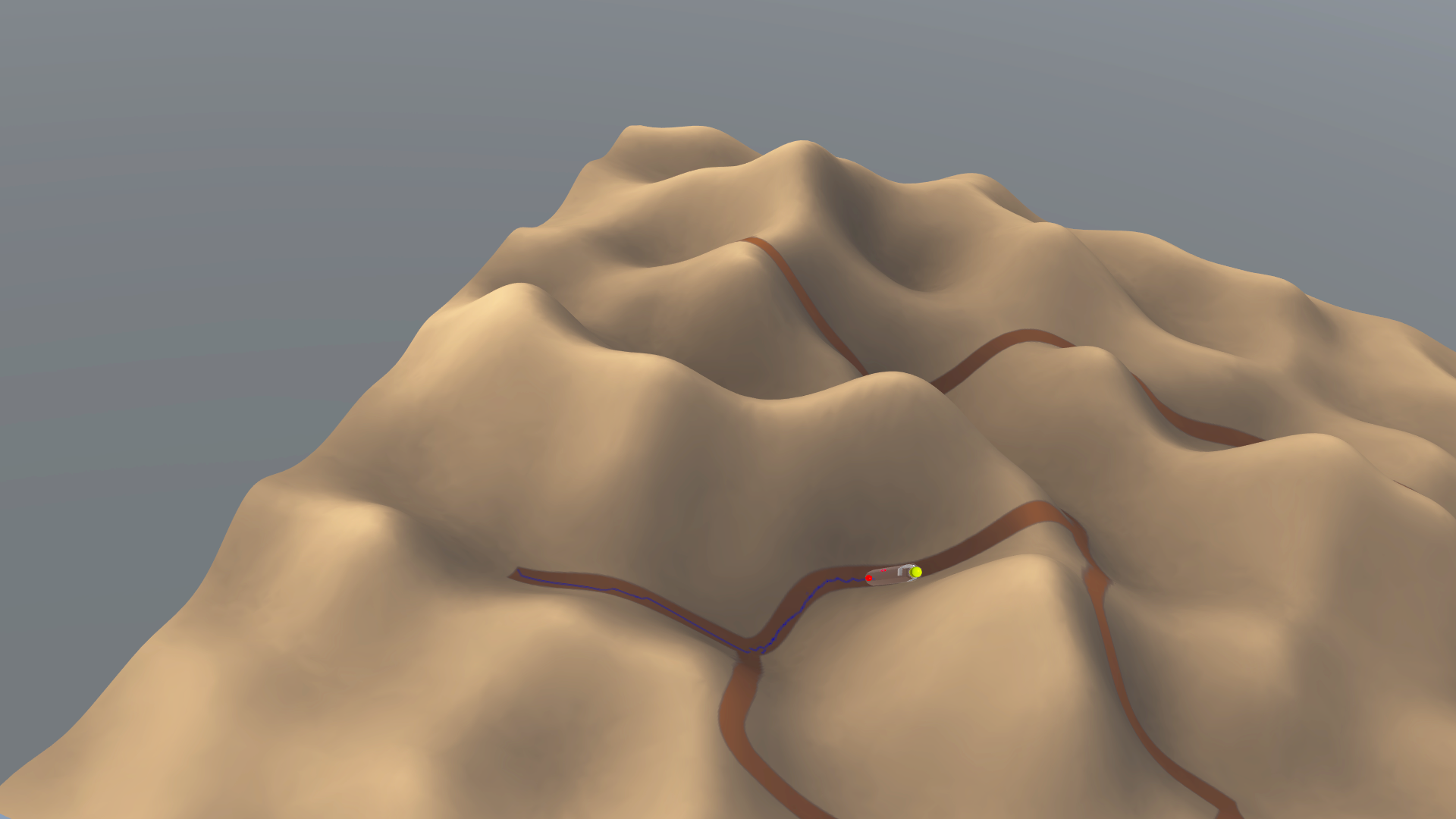}
  	\caption{Visually Marked Grooves Brushing}
  	\label{fig:groove}
  \end{subfigure}%
  \hfill
  \begin{subfigure}[b]{0.32\linewidth}
  	\centering
  	\includegraphics[width=\textwidth,trim={24cm 4cm 14cm 17cm},clip]{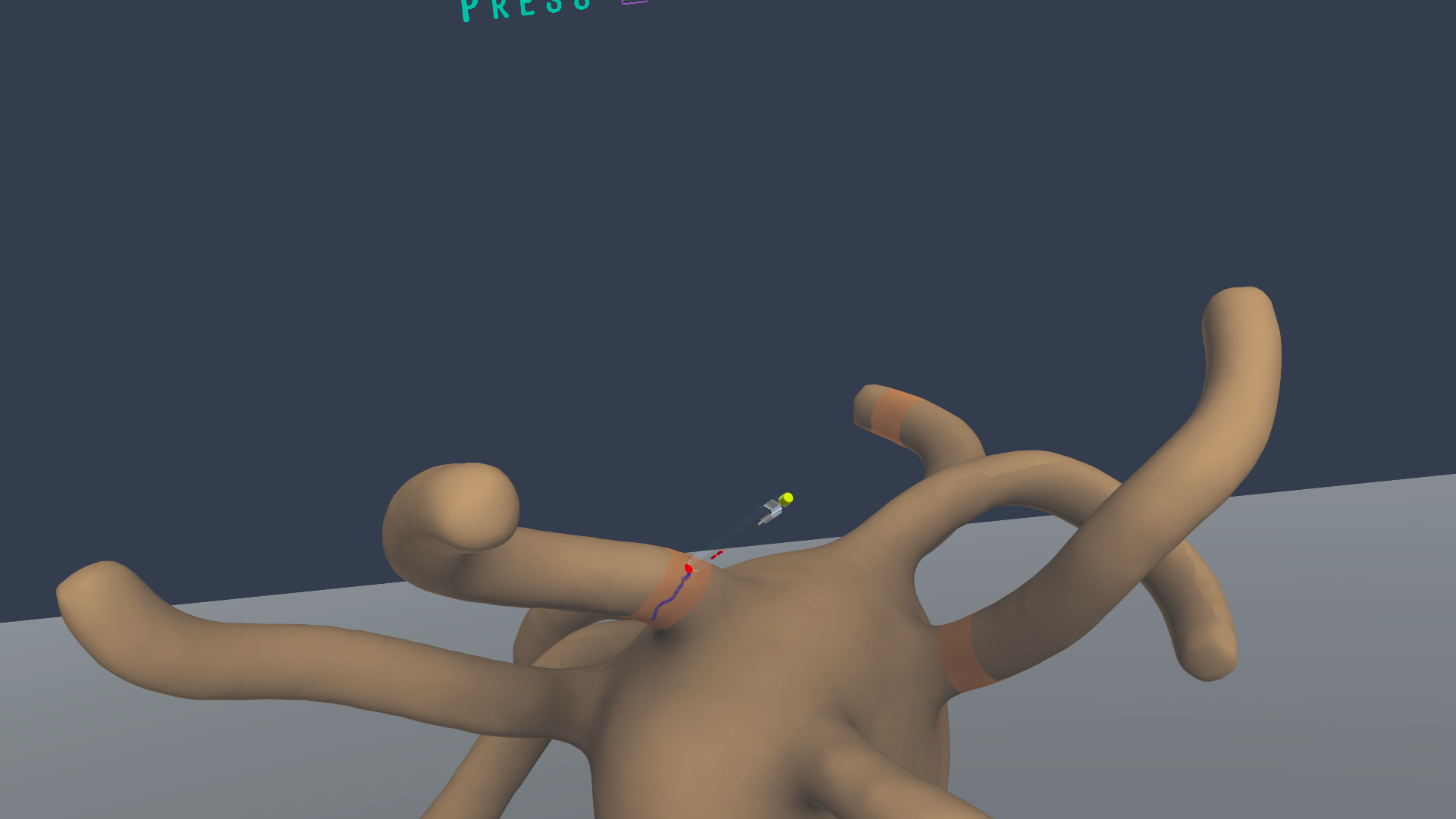}
  	\caption{Visually Annotated Regions Brushing}
  	\label{fig:annotation}
  \end{subfigure}%
  \subfigsCaption{All of the surface visualization interactions with the proxy for force-based haptic stylus.}
  \label{fig:inter_tasks}
\end{figure*}

\paragraph{\textbf{\texttt{Point Localization}}} entails selecting a specific point on the surface to draw meaningful information from \revisionb{either} inherent features of the surface or features indirectly mapped to \revisionb{it} via functions, textures or decals \cite{rocha_decal-maps_2017}, or a combination thereof. \revisionb{Based on our criteria, we identified three representative point localization interactions for surface visualizations.}

\emph{Protrusion Point Localization} entails localizing the highest protruded point on a surface against its base surface level. We procedurally generated surfaces (\cref{fig:protrusion}) with various protruded regions as described in~\cref{sec:surface_models}. The protrusion value at a point was computed by calculating its distance from the lowest point along the elevation axis of the surface and scaling it to a numeric range that represents realistic elevation data. 

\emph{Depression Point Localization} entails localizing the lowest depression point on a surface \revisionb{relative to} its base level. We sculpted a hemisphere-shaped surface model (\cref{fig:depression}) with various depression regions as described in Sec. \ref{sec:surface_models}. The surface depression value at a point was computed by calculating its distance from the centroid of the circle representing the hemispheric surface before deformation and scaling it to a realistic numeric range to assist interpretation. A region depressed \revisionb{relative to} its surroundings provides different visuo-haptic cues as compared to a protruded region. This task serves as a good benchmark for evaluating our three interaction modes on a surface \revisionb{with many} occluded regions.

\emph{Random Value Point Localization} involves finding the maximum function value displayed on the surface (\cref{fig:randval}). We created synthetic functions by combining 2D anisotropic Gaussians with random means and variances, visualized as contours with a diverging color map. Visual assistance is necessary because function values on the surface are independent of its shape. We used a color map with seven colors/contours, one representing the highest function value region. Finding the highest function value point still requires searching on the surface using the virtual stylus. The goal of this task is to evaluate how a strong visual cue affects the three interaction modes.

For additional visual assistance with the above tasks, we displayed the corresponding point value highlighted by the laser pointer on a floating widget directly above the surface visualization (see \cref{fig:inter_tasks}).

\paragraph{\textbf{\texttt{Curve Brushing}}}
is another basic interaction task associated with surface visualizations. Applications include feature or region demarcation and annotations for segmenting and clipping surfaces~\cite{besancon_state_2021}. \revisionb{This technique is also used for} sketching over an arbitrary surface or mid-air strokes \cite{elsayed_vrsketchpen_2020, jackson_lift-off_2016,keefe_drawing_2007,arora_experimental_2017}. Curves are brushed on the surface by pointing directly \revisionb{at it} using various input \revisionb{devices}, such as a mouse, tablet, stylus, or a handheld controller, and then moving along the features of interest. For our study, we identified three representative \texttt{Curve Brushing} interaction tasks for surface visualizations.


\emph{Visual Curve Brushing} involves brushing a curve within a visually highlighted band on the surface. We created procedural surfaces with landmarks for occlusions and clutter (see Sec. \ref{sec:surface_models} and \cref{fig:contour}). We generated curves using 2D modelling tools and offset them to create bands, which were then mapped to the surface as a texture. The modelled curves represent the medial axis of the bands and are independent of surface features. This task assesses snap-to-surface haptic interaction while brushing based on visual cues, regardless of surface topography changes, compared across haptic and no-haptic modes of interaction.


\emph{Visual Groove Brushing} involves brushing curves on visually highlighted grooves on a surface (\cref{fig:groove}). These grooves align with valleys formed by \revisionb{multiple protruded regions. We generated surfaces procedurally} (see Sec. \ref{sec:surface_models}) and manually highlighted the grooves with bands to aid visual querying and movement. When using the haptic stylus, the grooved regions guide the stylus to stay within them. This task examines whether snap-to-surface haptics offer advantages over haptic and no-haptic interactions when the surface indirectly provides haptic assistance.


\emph{Visual Annotation Brushing} involves annotating curves on the surface (\cref{fig:annotation}). Annotation regions highlight surface features, sourced from a catalog of object models (see Sec. \ref{sec:surface_models}). Models include a vase, cup, pliers, octopus, bear, and a plane, offering diverse interaction features. We manually highlighted features, mainly in regions with high occlusion, using bands to guide annotation. We aim to determine if snap-to-surface haptic interaction offers advantages over other modalities in regions with high occlusion but similar topography.

\revisionb{For all \texttt{Curve Brushing} tasks, we recorded the brushed curves by tagging the texels corresponding to the points highlighted by the interaction mode while brushing. Due to variable speed of haptic stylus, the pointer sometimes skipped texels during while brushing. To fix this, we linked the previous and current pointer positions via linear interpolation.}
\section{Evaluation}
\subsection{Research Questions}
Our user study focuses on addressing the following research questions.
\begin{enumerate}
[topsep=0pt,itemsep=0pt,partopsep=0pt, parsep=0pt]
    \item How do completion time and accuracy compare across immersive \texttt{Point Localization} tasks when using snap-to-surface haptic interaction (\texttt{HAPTIC\_SNAP}) versus the other modalities namely: haptic-only (\texttt{HAPTIC}) and no-haptic (\texttt{NO\_HAPTIC})?
    \item How do completion time and accuracy compare across immersive \texttt{Curve Brushing} tasks when utilizing snap-to-surface haptic interaction (\texttt{HAPTIC\_SNAP}) versus the other modalities namely: haptic-only (\texttt{HAPTIC}) and no-haptic (\texttt{NO\_HAPTIC})?
\end{enumerate}

\subsection{Study Design}
We followed a within-subject design for our user study, where participants were exposed to all three types of interaction modes to perform interaction tasks enumerated in~\cref{sec:interaction_design}.

\textbf{Task/Trials}: \revisionb{The \texttt{Point Localization} and \texttt{Curve Brushing} tasks outlined in~\cref{sec:interaction_design} are now referred to as \texttt{TaskProtrusion}, \texttt{TaskDepression} \texttt{TaskRandVal}, and \texttt{TaskCurve}, \texttt{TaskGroove}, \texttt{TaskAnnotation} respectively. To account for variability in surface visualizations, each of the six tasks was further divided into two trials with different levels of surface complexity. Each participant performed $36$ trials, divided into three sessions separated by at least $15$ minutes. For each session, participants performed $12$ trials, $2$ trials per task, while using a different interaction mode every session. The order of tasks, interaction mode and trials were counterbalanced using a balanced Latin square~\cite{mackenzie_chapter_2013}.}

\textbf{Practice Session}: \revisionb{To build familiarity with the interaction modes, each session included four practice tasks. The first task focused on acclimating participants to the haptic stylus, scene navigation, and controls using a procedural surface for \texttt{Point Localization} and \texttt{Curve Brushing}. Next, participants brushed a curve around a cylindrical surface, introducing them to all three interaction modes.} \revisionc{These training tasks ensured participants were familiar with the interaction modality and comfortable with controls before starting their tasks.}

\textbf{Datasets}: We generated the surface visualizations for each task using the methods outlined in~\cref{sec:design_choices}. The surface models were generated through Python scripts and Blender.

\begin{figure}[tb]
  \centering 
  \includegraphics[width=0.8\columnwidth, trim={0mm 110mm 0mm 90mm},clip]{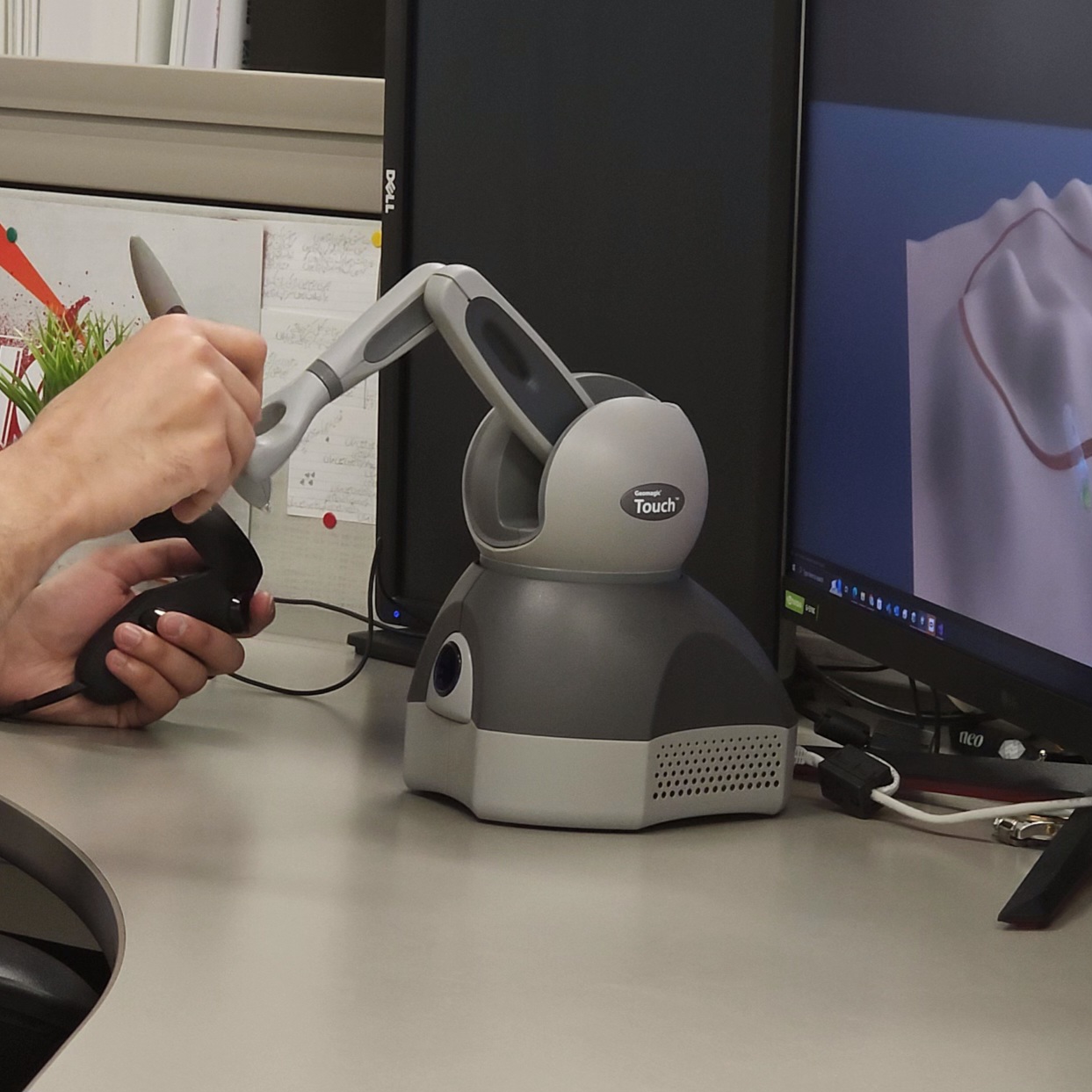}
  \caption{%
  	A right-handed participant using the force-based haptics device.
  }
  \label{fig:participant_haptic}
\end{figure}

\textbf{Apparatus}: The interaction modalities were implemented in Unity on a desktop computer (AMD Ryzen 7, 32 GB of RAM with a GeForce RTX3060 GPU) with an Oculus Rift S headset which provides a comfortable VR experience. We utilized the GeoMagic\textsuperscript{\sffamily\textregistered} Touch\textsuperscript{\sffamily\texttrademark} Professional haptic stylus to ensure realistic and accurate force-based haptic sensations. The sessions were conducted using a seated VR setup with the haptic stylus placed on the dominant hand side of the participants. Seating height was adjusted to allow armrest or table support, and the haptic device was aligned while the stylus was in its inkwell in-line with the direction of the participant’s arm. \cref{fig:participant_haptic} shows a picture of a participant using the haptic stylus for one of the trials.

\textbf{Participants}: We recruited a total of $24$ participants ($6$ females and $18$ males) by advertising the study to undergraduate and graduate students, faculty and postdoctoral fellows \revisionc{from various disciplines}. Age ranged from $20$ to $48$ (mean=$26.4$, SD=$5.16$), all the participants were right-handed. Prior experience using VR headsets in the context of 3D mapping applications such as Google Earth VR was used as a recruitment criterion. \revisionc{Participant's experience with VR was self-reported and ranged from novices to experts. They were screened through theVRISE test
~\cite{sharples_virtual_2008} (score < $25$) to avoid the effects of nausea and cybersickness in the study}. 
Prior experience with haptic devices was not necessary for participation, although $15$ participants reported having prior experience with them.

\subsection{Procedure}

Our study comprised three sessions, each with two phases: a practice session and main trials. Participants typically spent 30 to 45 minutes completing a session, followed by a break of at least 15 minutes to minimize fatigue. In the first session, we introduced the study's purpose, then led participants through a training session to familiarize them with all three interaction modes (\texttt{NO\_HAPTIC}, \texttt{HAPTIC}, and \texttt{HAPTIC\_SNAP}), and the tasks. They were encouraged to ask questions and take time to adjust to the interaction modality and controls before beginning the main trials. 

In each trial, participants followed on-screen instructions to perform the designated tasks and submitted their results when they believed they had completed the tasks. They were instructed to complete the trials with both speed and accuracy. Using a VR handheld controller in their non-dominant hand, they navigated the visualization, while performing point localization or curve brushing interactions with their dominant hand, employing the assigned modality. To proceed, they pressed the joystick button on their non-dominant hand controller.

Finally, following the last session, we conducted a post-study questionnaire to gather feedback on the visualizations and interaction modes used in the study. This questionnaire included multiple-choice questions based on the NASA Task Load Index~\cite{nasa} and provided participants with an opportunity to offer suggestions or comments.

\subsection{Measures}
\label{sec:measures}

We collected two main objective measures per trial: \emph{task completion time}, tracking from trial start to submission, and \emph{user responses}, including localized point values and brushed curve textures for \texttt{Point Localization} and \texttt{Curve Brushing} tasks, respectively. Additionally, per-frame logging included stylus, head-mounted display (HMD), and handheld VR controller positions, interaction mode, navigation transformations, touch events (excluding \texttt{NO\_HAPTIC}), and selection, draw, and erase events. Brushed curves were stored as textures recording the UV coordinates of the brushed curves. Participants' trials were randomly videotaped and screen-recorded.

\section{Results}
\label{sec:results}

We analyze the performance of \texttt{NO\_HAPTIC}, \texttt{HAPTIC}, and \texttt{HAPTIC\_SNAP} interaction modes using estimation techniques applied to data from 720 trials. Following recommended practice~\cite{cumming_inference_2009,dragicevic_fair_2016}, we report pairwise differences between means with $95\%$ confidence intervals (CI) and avoid p-value statistics,
though they can be derived from the data. 
Confidence intervals are constructed using empirical bootstrapping. We begin with an overview of each interaction modality's performance before diving into a more detailed analysis.

\subsection{Overview}
\begin{figure}[t]
  \centering 
  \includegraphics[width=\columnwidth, trim={3mm 6mm 3mm 7mm},clip]{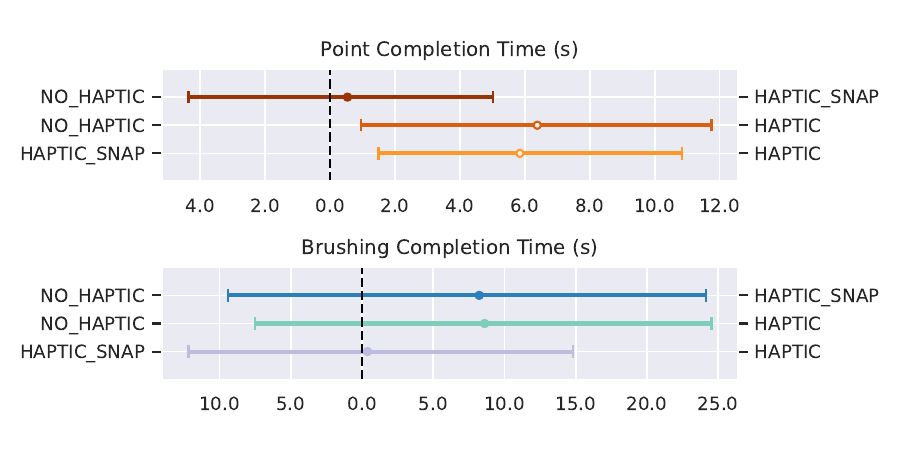}
  \caption{%
  A holistic comparison between \textcolor{point_color}{\texttt{Point Localization}} and \textcolor{brush_color}{\texttt{Curve Brushing}} tasks with respect to total time taken (mean) for completion. Significant differences are highlighted with a \textcolor{sig_color}{white} marker. 
  }
  \label{fig:completion_time_overall}
\end{figure}

\begin{figure}[t]
  \centering 
  \includegraphics[width=\columnwidth, trim={3mm 6mm 3mm 7mm},clip]{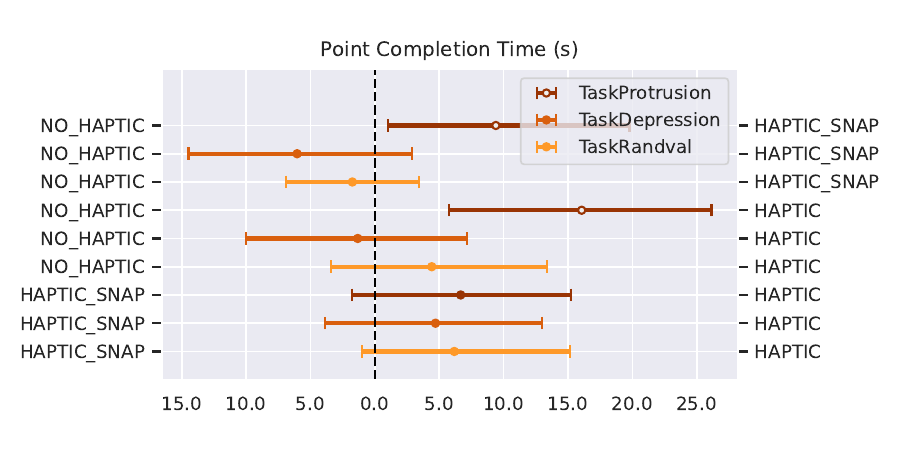}
  \caption{%
  Per-task difference of completion time for \textcolor{point_color}{\texttt{Point Localization}}.
  }
  \label{fig:point_completion_time}
\end{figure}

For \texttt{Point Localization} tasks, the \texttt{HAPTIC} mode took $6.38$s (mean) (CI = $[0.95,\,11.75]$) longer compared to the \texttt{NO\_HAPTIC} mode, and $5.84$s (CI=$[1.49,\,10.82]$) longer compared to the \texttt{HAPTIC\_SNAP} mode. The \texttt{NO\_HAPTIC} and \texttt{HAPTIC\_SNAP} modes had no significant difference (\cref{fig:completion_time_overall} top).
Additionally, we found no significant difference between the modes for \texttt{Curve Brushing} tasks (\cref{fig:completion_time_overall} bottom). On a more granular level, for \texttt{Point Localization} tasks, the on-surface interaction modes, \texttt{HAPTIC\_SNAP} and \texttt{HAPTIC}, took $9.39$s (CI=$[1.0,\,19.77]$) and $16.06$s (CI=$[5.77,\,26.14]$) more than \texttt{NO\_HAPTIC} respectively for \texttt{TaskProtrusion} (\cref{fig:point_completion_time}). Our results suggest that other task-wise differences for \texttt{Point Localization} were not significant and more data is needed to get significant outcomes. Similarly, \texttt{Curve Brushing} tasks showed no task-wise significant differences in completion time. 

\begin{figure}[t]
  \centering 
  \includegraphics[width=\columnwidth, trim={3mm 6mm 3mm 7mm},clip]{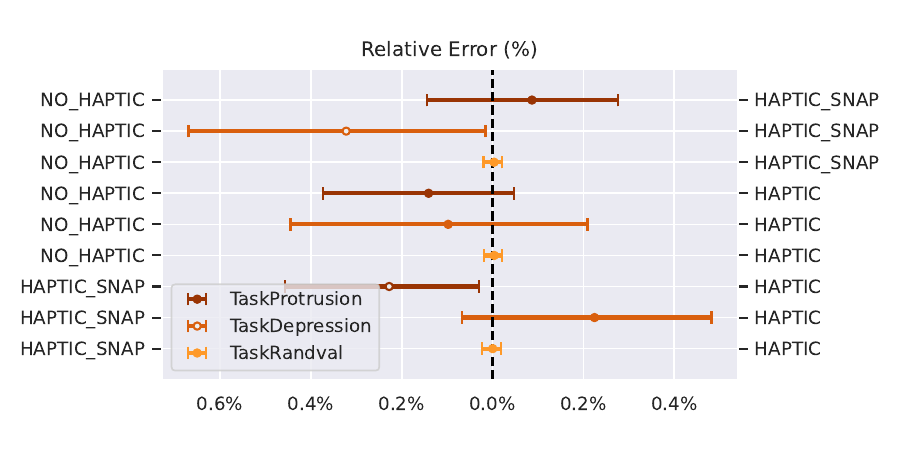}
  \caption{%
  Comparison of relative error for \texttt{\pointc{Point Localization}} tasks that incurred a relative error less than $5\%$.
  }
  \label{fig:point_error_per_task}
\end{figure}

We now report the errors made in \texttt{Point Localization} and \texttt{Curve Brushing} tasks for the three interaction modes. For \texttt{Point Localization}, we computed the relative error as compared to the ground-truth localization point. The relative errors were largely concentrated near zero. Therefore, for our bootstrapping analysis, we first filtered the errors and retained only those that were within a $5\%$ error range from the ground-truth. At a per-task level, we observe that the error was higher for \texttt{NO\_HAPTIC} compared to \texttt{HAPTIC\_SNAP} in \texttt{TaskDperession} ($0.32\%$, $[0.01\%,0.66\%]$), whereas the error for \texttt{HAPTIC\_SNAP} was higher compared to \texttt{HAPTIC} for \texttt{TaskProtrusion} ($0.22\%$, $[0.02\%,0.45\%]$). For \texttt{TaskDepression}, a skewness towards the \texttt{HAPTIC} mode indicates lower error for \texttt{HAPTIC\_SNAP}, but more data is needed for a conclusive outcome. Lastly, we observed no differences for \texttt{TaskRandval} between any interaction modes (\cref{fig:point_error_per_task}). 

For \texttt{Curve Brushing} tasks, we quantified the deviation of the brushed curve from the medial axis of the visualized band that the participants were asked to follow. We first computed the Euclidean distance transform (EDT) on the brushed texture image and compared it to the EDT of the medial axis of the band texture shown on the visualization. Finally, the following root mean square error (RMSE) was computed to quantify curve deviation.

\vspace{-11px}
\begin{equation}
\Bigl[
\frac{1}{\lvert \mathcal{B} \rvert}
\sum_{(i,j) \in \mathcal{B} } 
\bigl(
D_{\text{G}}[i,j] - D_{\text{X}}[i,j]
\bigr)^2
\Bigr]^{\tfrac{1}{2}}.
\label{eqn:curve_deviation}
\end{equation}
\vspace{-10px}

\begin{figure}[t]
  \centering 
  \includegraphics[width=\columnwidth, trim={3mm 5mm 3mm 7mm},clip]{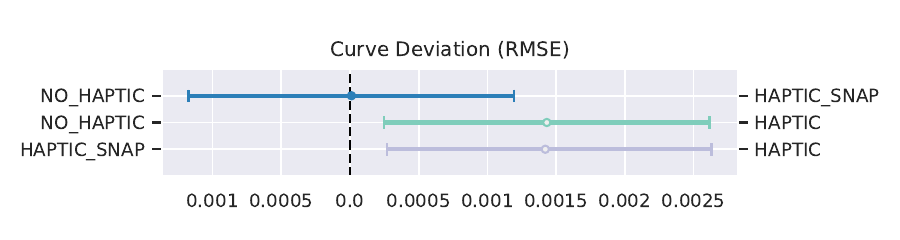}
  \caption{%
  Comparison of curve deviation for \texttt{\brushc{Curve Brushing}} tasks. 
  }
  \label{fig:curve_deviation_overall}
\end{figure}

\begin{figure}[t]
  \centering 
  \includegraphics[width=\columnwidth, trim={3mm 6mm 3mm 7mm},clip]{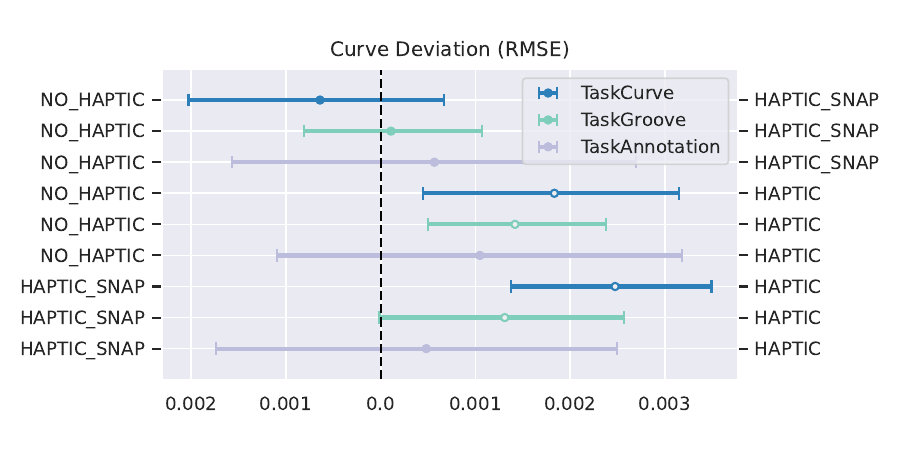}
  \caption{%
  Task-wise comparison of curve deviation for \texttt{\brushc{Curve Brushing}}.
  }
  \label{fig:curve_deviation_per_task}
\end{figure}

\noindent $D_\text{G}$ is the EDT of the reference medial axis texture, $D_\text{X}$ is the EDT of the brushed texture ($\text{X} \in \{$\texttt{HAPTIC\_SNAP}, \texttt{HAPTIC}, \texttt{NO\_HAPTIC}$\}$), and $\mathcal{B}$ is the set of texels that fall within the band shown to the participants. Holistically, we observe that the curve deviation when using \texttt{HAPTIC} was higher than \texttt{NO\_HAPTIC} ($0.0014$, $[0.0002,0.0026]$) and \texttt{HAPTIC\_SNAP} ($0.0014$, $[0.0026,0.0026]$) modes, whereas the difference between \texttt{NO\_HAPTIC} and \texttt{HAPTIC\_SNAP} was not significant 
(\cref{fig:curve_deviation_overall}). On a granular level, we see that \texttt{HAPTIC} had higher error compared to \texttt{NO\_HAPTIC} and \texttt{HAPTIC\_SNAP} for \texttt{TaskCurve} (($0.0018$, $[0.00044, 0.0031]$) and  ($0.0024$, $[0.0013, 0.0034]$)) and \texttt{TaskGroove} (($0.0014$, $[0.00049s, 0.0023s]$) and ($0.0013$, $[0.0000134, 0.002]$)), but \texttt{TaskAnnotation} had no significant differences~(\cref{fig:curve_deviation_per_task}). 

\subsection{Point Localization}
\begin{figure}[t]
  \centering 
  \includegraphics[width=0.75\columnwidth, trim={3mm 6mm 3mm 10.mm},clip]{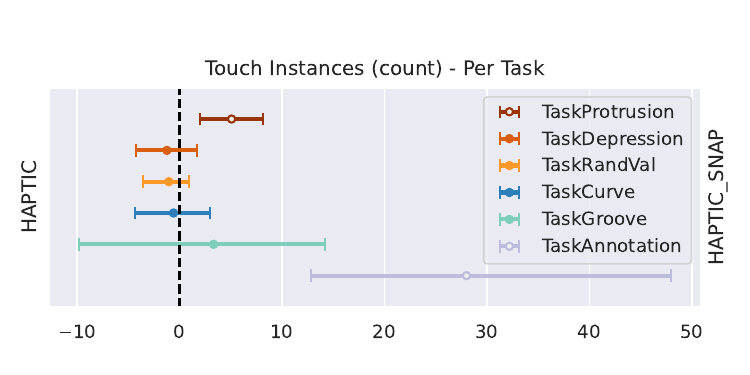}
  \caption{%
  A difference between the number of touch instances for every \pointc{Point Localization} and \brushc{Curve Brushing} tasks. 
  }
  \label{fig:per_task_touch_count}
\end{figure}

\revisionb{\texttt{HAPTIC\_SNAP} is faster compared to the \texttt{HAPTIC} mode, but the difference is not significant at a per-task level. Moreover, its accuracy was worse in \texttt{TaskProtrusion}. Participants found it harder to localize a point on protruded regions with steeper slopes using \texttt{HAPTIC\_SNAP}, as the snapping force caused the stylus to slide to the side of the slope. This led participants to touch the surface more frequently compared to \texttt{HAPTIC} ($5.08$, $[1.99,\,8.16]$) for localization (\cref{fig:per_task_touch_count}). When contrasted with \texttt{NO\_HAPTIC}, both on-surface modes, \texttt{HAPTIC} and \texttt{HAPTIC\_SNAP}, were slower in \texttt{TaskProtrusion}; we also found no significant difference in selection accuracy.} This is mainly because participants could point to the protruded region from afar with the stylus relying solely on visual cues from the surface, making the localization process faster while providing enough stability to select points accurately. 

\begin{figure}[t]
  \centering 
  \includegraphics[width=\columnwidth, trim={3mm 6mm 3mm 7mm},clip]{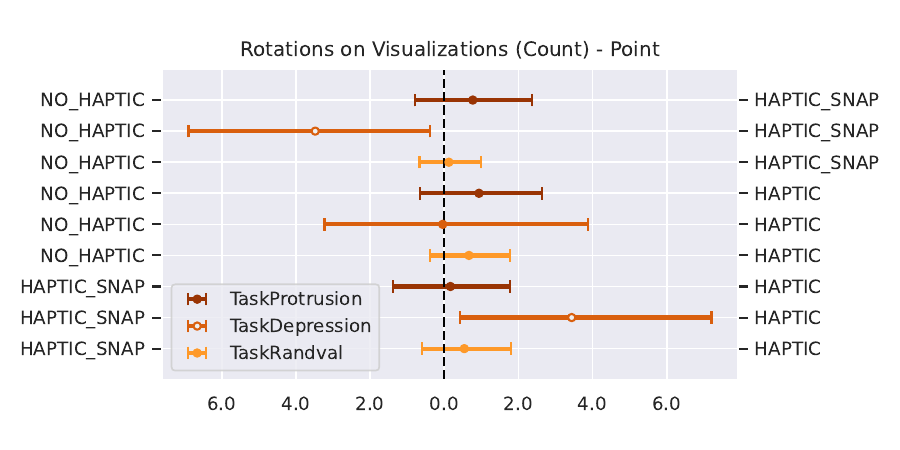}
  \caption{%
  The difference in the number of rotations performed for \pointc{Point Localization} tasks.
  }
  \label{fig:point_task_rotations}
\end{figure}

When observing \texttt{TaskDepression}, we find that \texttt{HAPTIC\_SNAP} outperforms \texttt{NO\_HAPTIC} in accuracy; it also seems to have better accuracy compared to \texttt{HAPTIC} but the result is not significant (\cref{fig:point_error_per_task}). Furthermore, \texttt{HAPTIC\_SNAP} took fewer rotations to localize the point compared to \texttt{NO\_HAPTIC} ($3.47$, $[0.39,6.89]$) and \texttt{HAPTIC} ($3.43$, $[0.43,7.20]$) modes (\cref{fig:point_task_rotations}). This underscores that on a surface with significant occlusion and challenging depth perception, the snapping feature of the stylus enabled participants to effortlessly navigate to the area of interest for selection. This behavior contrasts with the findings of \texttt{TaskProtrusion}, where the haptic stylus struggled with localization on protruded regions but performed better in depressed regions.


For \texttt{TaskRandVal}, no significant difference in time or accuracy is observed across the modes due to the strong reliance on visual cues from relatively flat surfaces with distinct contours for localizing function values, where the haptic feedback did not significantly impact the outcomes.

\subsection{Curve Brushing}
\begin{figure}[t!]
  \centering 
  \includegraphics[width=\linewidth]{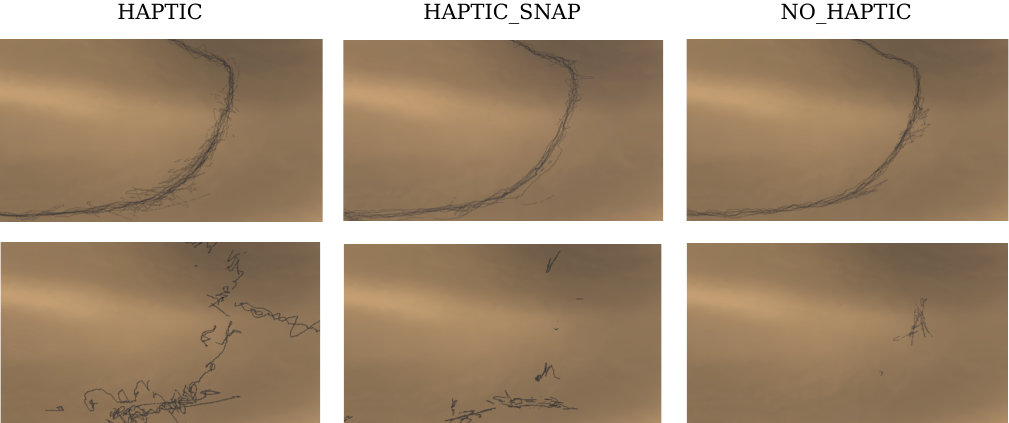}
  \caption{Heatmap of curves brushed (top) and corrections made (bottom).
  }
  \label{fig:curves_heatmap}
\end{figure}

We observe no difference in the time taken by participants to brush curves. However, curves deviated more from the medial axis for \texttt{HAPTIC} compared to \texttt{NO\_HAPTIC} and \texttt{HAPTIC\_SNAP}.
\revisionb{During the trials, we instructed participants to stay within the visualized band and make corrections when the curve left the bounds of the band. They were also asked to brush the curves as close to the center of the band as possible. Upon analyzing the curves, we observed that when the stylus was angled away from the normal, the normal force exerted by the \texttt{HAPTIC} stylus tended to push it away from the medial axis. This required participants to exert a counterbalancing force to stay on the intended path, as illustrated in \cref{fig:curves_heatmap}. In these regions, \texttt{HAPTIC} participants frequently made corrections. Conversely, \texttt{HAPTIC\_SNAP} and \texttt{NO\_HAPTIC} showed slightly improved performance under similar conditions. However, the rapidly changing topography of the surface visualization affected brushing tasks for both \texttt{HAPTIC\_SNAP} and \texttt{NO\_HAPTIC}; minor directional changes to the stylus led to mistakes on highly varying surfaces. }

We also noticed a remarkable difference in the regularity (smoothness) of the curves drawn using the interaction modalities. To further quantify this, we computed the EDT on the brushed texture image, followed by a Laplacian of the EDT. Finally, we computed the standard deviation (SD) of the Laplacian to characterize the irregularity of the curve. Specifically, the Curve Irregularity is defined as
\vspace{-5px}
\begin{equation}
\mathrm{{SD}} \bigl[
\left\{\nabla^2{D_X}[i,j] : (i,j) \in \mathcal{B} 
\right\}
\bigr].
\label{eqn:curve_irregularity}
\end{equation}
\vspace{-15px}

\begin{figure}[t]
  \centering 
  \includegraphics[width=\columnwidth, trim={3mm 5mm 3mm 7mm},clip]{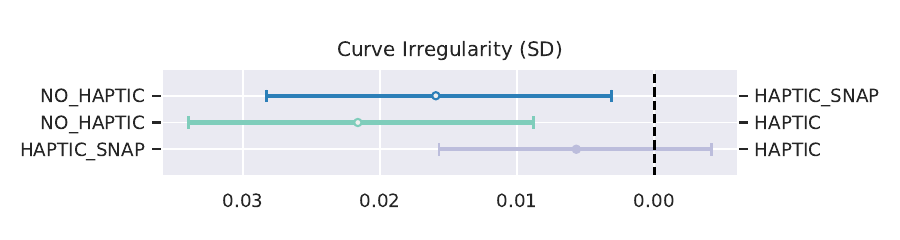}
  \caption{%
  Overall comparison of curve irregularity for \brushc{Curve Brushing} tasks.
  }
  \label{fig:curve_irregularity_overall}
\end{figure}

\begin{figure}[t]
  \centering 
  \includegraphics[width=\columnwidth, trim={3mm 6mm 3mm 7mm},clip]{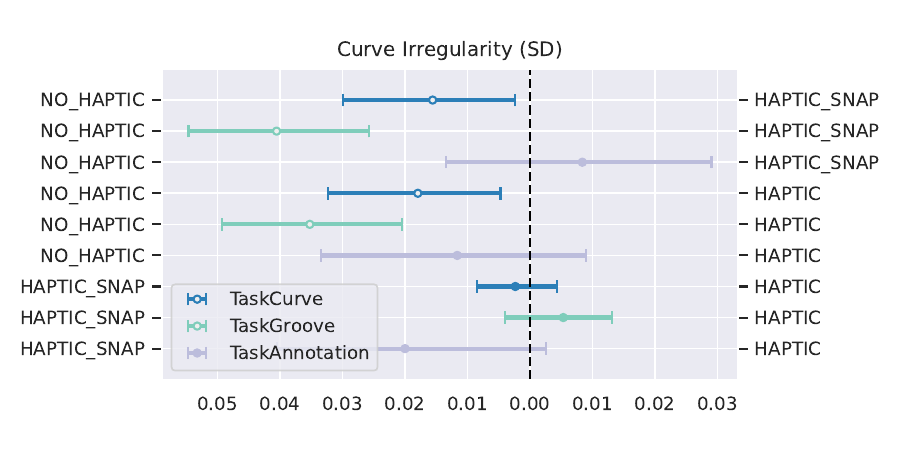}
  \caption{%
  Task-wise comparison of curve irregularity for \texttt{\brushc{Curve Brushing}}.
  }
  \label{fig:curve_irregularity_per_task}
\end{figure}

\begin{figure}[t!]
  \centering 
  \includegraphics[width=\linewidth, trim={58.5cm 23cm 4.2cm 20.5cm}, clip]{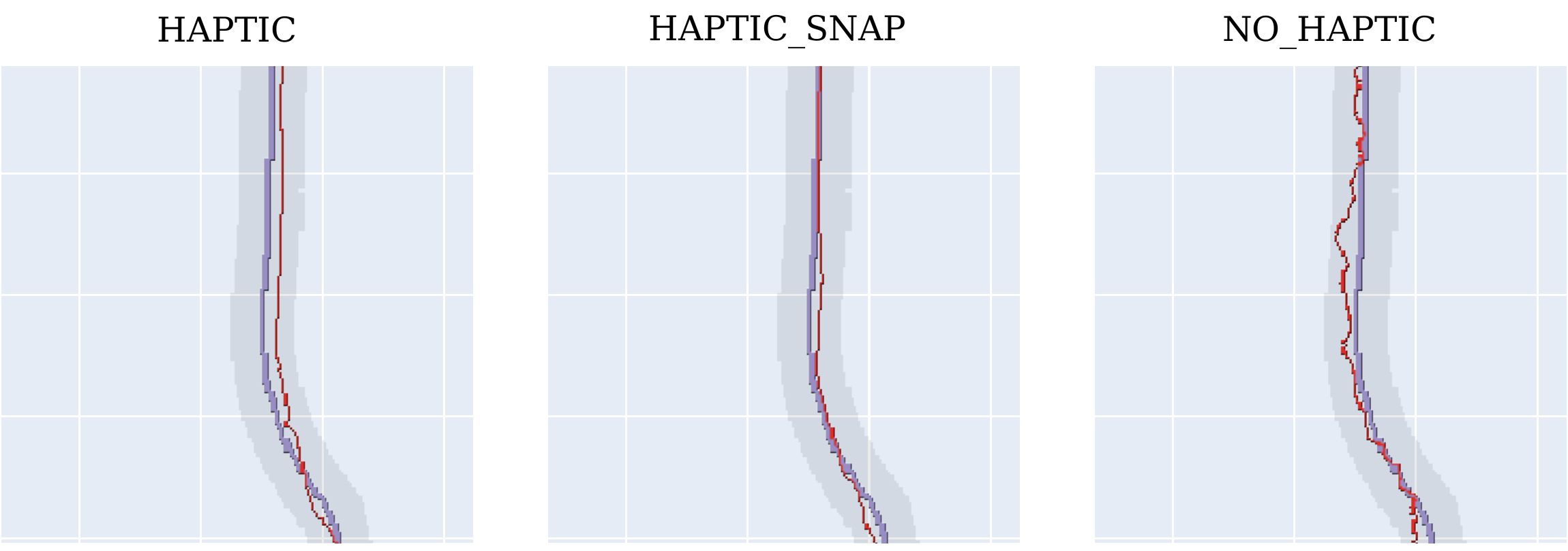} 
  \caption{A comparison of curves brushed using different haptic modes.}
  \label{fig:curve_smoothness}
\end{figure}

\revisionb{\noindent It measures jitters in the curves attributable to hand stability when brushing on surfaces (higher values indicate higher irregularity). With \texttt{NO\_HAPTIC}, participants brushed curves that were significantly more irregular than with \texttt{HAPTIC} ($0.021$, $[0.008, 0.03]$) and \texttt{HAPTIC\_SNAP} ($0.015$, $[0.003, 0.02]$), while \texttt{HAPTIC} and \texttt{HAPTIC\_SNAP} showed no significant difference (\cref{fig:curve_irregularity_overall}). This trend continues on a per task level, as \texttt{NO\_HAPTIC} had irregular curves compared to other modes in \texttt{TaskCurve} and \texttt{TaskGroove}, but \texttt{TaskAnnotation} saw no significant difference (\cref{fig:curve_irregularity_per_task}). \cref{fig:curve_smoothness} visually depicts the irregularity of curves drawn by \texttt{HAPTIC}, \texttt{HAPTIC\_SNAP} and \texttt{NO\_HAPTIC}. The \texttt{HAPTIC} curves were smoother but deviated from the medial axis, while \texttt{NO\_HAPTIC} curves were closer to the medial axis, but more irregular. Curves brushed with \texttt{HAPTIC\_SNAP} were both smooth and accurate.}

\begin{figure}[t!]
  \centering 
  \includegraphics[width=\linewidth, trim={3mm 2mm 47mm 2mm},clip]{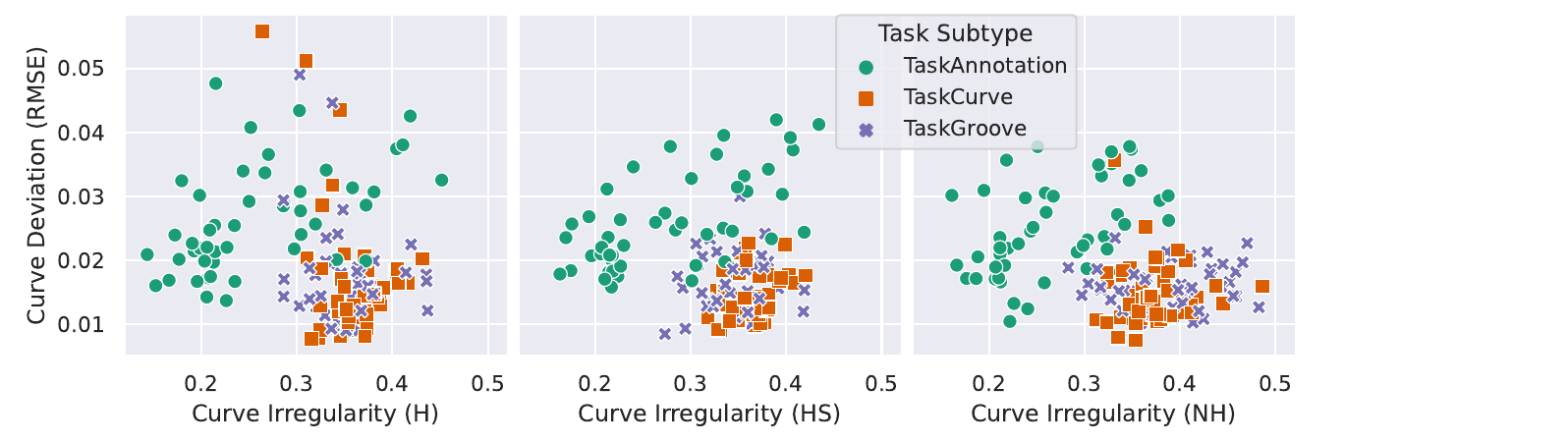}
  \caption{Scatter plot of curve irregularity and deviation from the medial axis.
  }
  \label{fig:brushing_correlations}
\end{figure}

The relationship between curve deviation and irregularity is shown in~\cref{fig:brushing_correlations}. In \texttt{TaskAnnotation}, all modes exhibit lower irregularity, but \texttt{HAPTIC} shows wider deviation, possibly due to complex surface features. For \texttt{TaskCurve} and \texttt{TaskGroove}, \texttt{NO\_HAPTIC} has higher irregularity but similar deviation to \texttt{HAPTIC\_SNAP}. While \texttt{HAPTIC} has similar irregularities to \texttt{HAPTIC\_SNAP}, it shows greater deviation. \texttt{HAPTIC\_SNAP} yields smoother curves, suggesting its snapping force reduces the need for firm pressure, resulting in smoother and less error-prone curves. Across all modes, higher irregularity correlates with greater deviation.


\revisionb{Inspecting the transformations for \texttt{Curve Brushing} tasks, we observe no significant difference in the number of translation and rotation events at both aggregated and per-task levels. However, participants touched the surface more often to draw curves using \texttt{HAPTIC\_SNAP} compared to \texttt{HAPTIC} ($28.0$, $[12.7,47.9]$) (\cref{fig:per_task_touch_count}). This indicates that participants using \texttt{HAPTIC\_SNAP} faced difficulties in drawing curves on complex surface regions with dense small-scale features, as the snapping force made navigation more challenging.}



    
\section{Participant Feedback}

\begin{figure}[t]
  \centering 
  \includegraphics[width=\columnwidth, trim={0cm 0.5cm 0cm 1cm}, clip]{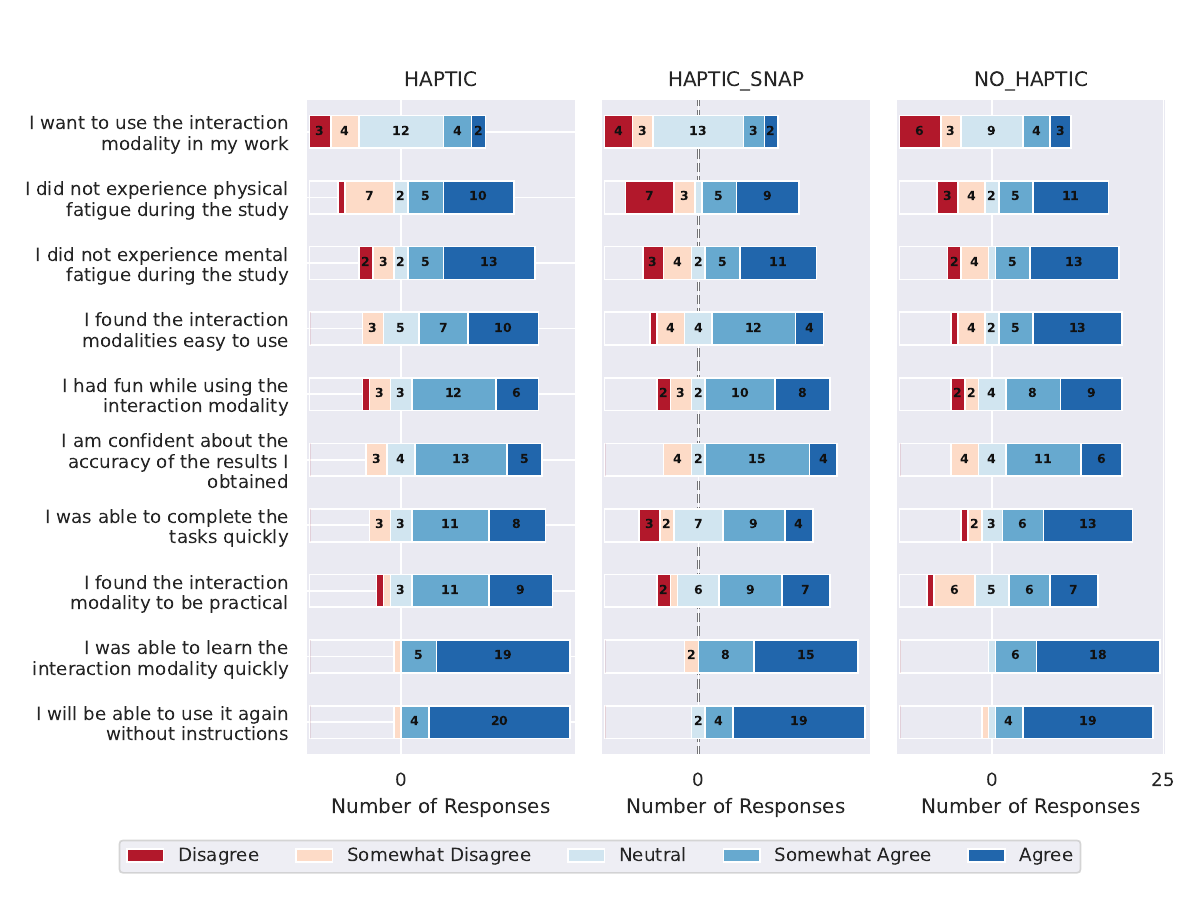}
  \caption{Likert scale summarizing participant responses.
  }
  \label{fig:participants_response}
\end{figure}

\revisionc{The post-study questionnaire results, shown in \cref{fig:participants_response}, indicate that \texttt{HAPTIC\_SNAP} users reported higher mental and physical fatigue due to constant feedback from the force-based stylus, aligning with Choi \textit{et al.}~\cite{choi_force_2005}. Despite this fatigue, participants achieved smoother and more accurate curves. Although \texttt{HAPTIC\_SNAP} felt time-consuming, no significant time difference was noted (\cref{sec:results}). \texttt{NO\_HAPTIC} was seen as less practical, but users were confident in their results, showing accuracy despite reduced smoothness. Participants adapted quickly, enjoyed the experience, and felt prepared for future use without guidance. Overall satisfaction scores (out of 5) were: \texttt{HAPTIC} 3.84, \texttt{NO\_HAPTIC} 3.64, and \texttt{HAPTIC\_SNAP} 3.48. This suggests \texttt{HAPTIC} mode was easier to use, while \texttt{HAPTIC\_SNAP}, despite better performance, was rated lower, supporting claims about continuous force feedback.}

\revisionc{Participants provided feedback on interaction modes, emphasizing \texttt{HAPTIC\_SNAP}. They noted its accuracy and ease, which \textit{made tracing tasks easier} and \textit{improved accuracy} (P7). Curve tracing was simpler, and participants \textit{appreciated the reduced need for pressure when drawing curves} (P16). However, challenges included \texttt{HAPTIC\_SNAP} \textit{sometimes snapping to the wrong area or jumping between adjacent surfaces} (P12). Fatigue was also a concern as certain angles required \textit{uncomfortable arm positions, leading to tiredness} (P21).}

\revisionc{The \texttt{HAPTIC} mode was preferred for its ease of use and intuitive feel. Participants described it as \textit{"the best overall" for completing tasks effortlessly} (P6). It felt like using a real pen, \textit{providing a clear experience akin to touching a real object} (P12). This mode also resulted in \textit{less fatigue due to the freedom of movement} (P24). However, challenges included applying excessive force, leading to errors, especially on sloped surfaces where \textit{too much pressure caused the pen to slip} (P2). Some participants found \texttt{HAPTIC\_SNAP} \textit{better for maintaining accuracy} in these situations (P2). Additionally, \textit{cognitive load increased} with \texttt{HAPTIC} as participants \textit{had to guide themselves more}, and variable force feedback on changing topography made tracking difficult (P17). Sliding issues were also noted, \textit{with the pen easily sliding down hills during tracing tasks} (P22).}

\revisionc{In \texttt{NO\_HAPTIC} mode, participants appreciated reduced physical fatigue and flexibility, \textit{noting the absence of hand tiredness} (P1) and \textit{allowing more freedom} (P6). However, \textit{lower hand pressure led to reduced accuracy} (P13). Without haptic feedback, users experienced fatigue and imprecise curves, \textit{comparing it to using a laser pointer} (P2), which made \textit{staying close to the geometry difficult}, resulting in inaccurate drawings and no hand support (P17). Spatial awareness was problematic, making it \textit{hard to tell if tool touched the intended surface} (P17), and the lack of tactile feedback \textit{reduced immersion}, affecting overall experience (P23).}

\revisionc{Overall, participants suggested mode switching to improve accuracy, expressing the desire to \textit{alternate between modes} (P8). They acknowledged a significant learning curve linked with these modes, but felt it was manageable with practice, describing it as \textit{a novel and interesting way to interact} (P14). However, device limitations, notably in axis control, were posed as a challenge. Participants found it \textit{difficult to move in a straight line left or right and compensated by rotating the device} (P5).}

\section{Takeaways}
In this section, we summarize our findings as takeaways to provide design guidelines for future researchers using force-based haptics for surface visualization interactions. 

\textbf{Takeaway\#1:} \textit{"Force-based haptics with snapping force excels in localizing depression points but struggles with protruded regions}. \revisionb{Our analysis and participant feedback (P7) indicate that users rely on snap force for precision in depressions. However, controlling snap force for accurate localization in protruded areas remains challenging.}

\textbf{Takeaway\#2:} \textit{"The accuracy and speed of localizing points using visual cues on surfaces are unaffected by the choice of interaction mode."} Our evaluation shows that when participants rely on distinct visual cues from the surface, their performance is consistent across all three modes.

\textbf{Takeaway\#3:} \revisionb{\textit{"Accuracy of curves brushed with haptic feedback is influenced by the angle of applied force relative to the surface normal."} Findings and feedback (P22, P12) indicate that participants using the haptic stylus produced smoother curves, but applying force at angles deviating from the surface normal caused divergence from the intended path.}


\textbf{Takeaway\#4:} \revisionb{\textit{"On-surface force feedback modes for brushing yields smoother curves."} Analysis and feedback (P6, P23) reveal that snapping reduces the user-exerted force toward the surface and enables tangential force application, assisting users in drawing smooth and accurate curves.}


\textbf{Takeaway\#5:} \textit{"Brushing on surfaces with dense, small scale features is challenging with the snapping force."} Analysis and feedback (P10) indicate that using the snapping force with the haptic device resulted in more surface contacts, leading to frustration in task completion.

\revisionc{\textbf{Takeaway\#6:} \textit{"Haptic and no-haptic modes can be interchanged during tasks."} We propose using a switch to activate assisted haptic mode on angled or bumpy surfaces, while Haptic-only mode suits narrow or protruded areas and no-haptics for tasks relying on visual cues.}

\section{Conclusion, Limitations and Future Work}

We explored the efficacy of on-surface interaction modes --- \textit{both assisted and unassisted}--- alongside a no-haptic mode, utilizing force-based haptics for surface visualizations. Our objective was to assess these modes' performance in visualization tasks complicated by occlusion and dependent on rendering quality for accurate interpretation. We conducted a user study focusing on common surface visualization interactions: point localization and curve brushing. Our findings reveal that haptic assistance, particularly snapping force, enhances the speed of occluded point localization (\texttt{TaskDepression}) but complicates localization in protruded areas (\texttt{TaskProtrusion}). Additionally, this assistance leads to smoother and more accurate curves, whereas unassisted interactions tend to stray from the target. Participant feedback was synthesized to underscore key insights, offering guidance for future research in this domain.

\revisionb{We acknowledge limitations in our study and suggest future research directions. On-surface haptics with snap assistance improved curve smoothness and accuracy by helping participants use snapping force to stay on the surface, reducing penetration force. Analyzing exerted force could offer more insights. Although some participants had haptic device experience, the interaction modes presented a significant learning curve. Future research could examine long-term effects on performance in a longitudinal study. While some tasks showed significant differences between modalities, others were nearly significant.} Further analysis with a larger\revisionc{, more diverse participant group with varied VR experience} could yield more definitive conclusions. \revisionc{Although we made efforts to reduce VR headset discomfort and fatigue, future work should further explore these effects to better understand interactions in VR environments.}


\revisionb{While our study used a force-based haptic device with a stylus design, which can be generalized to other similar devices,
it did not evaluate other types such as haptic gloves and advanced devices. Evaluating these devices is a topic of future research. We used a mix of procedurally generated explicit surface models; we believe that these surface models represent the variety of surfaces that users are likely to encounter}. \revisionc{Future research could explore implicit and genus 1+ surfaces with narrow holes, as these complex models may challenge the snap-assisted mode by requiring variable snap distances and force profiles}. We limited our study to point localization and curve brushing tasks. Future work can explore a broader range of haptic tasks such as multivariate data encoding on surfaces and interactions with deformable surfaces and volumes.

\revisionb{Our study focused only on the force-based stimulus from the haptic device which provides a physical feel of the surface's shape and depth. Future work could explore forms of force-based haptifications such as force vibrations and surface friction, to relay multiple data encodings in the form of multivariate surface visualizations. Our study focuses on one force profile; future research should investigate various profiles for different tasks. Feedback from participants} highlighted the potential benefits of switching between interaction modes, as suggested by existing studies \cite{choi_force_2005}, indicating the potential for developing an automated mode selection system for surface interactions.

\bibliographystyle{abbrv-doi-hyperref-narrow}

\bibliography{template, HapticsResearch}

\begin{thebibliography}{10}
\renewcommand*{\sfdefault}{PTSansNarrow-TLF}

\bibitem{geomagic}
{3D Systems}.
\newblock {Geomagic Touch User Guide}.
\newblock \url{https://support.3dsystems.com/s/article/Ethernet-Geomagic-Touch-Haptic-Device-User-Guide}.
\newblock Accessed: Jun 30, 2024.

\bibitem{nasa}
{AHRQ}.
\newblock {NASA Task Load Index | Digital Healthcare Research}.
\newblock \url{https://humansystems.arc.nasa.gov/groups/tlx/downloads/TLXScale.pdf}.
\newblock Accessed: Jun 30, 2024.

\bibitem{alabi_comparative_2012}
O.~S. Alabi, X.~Wu, J.~M. Harter, M.~Phadke, L.~Pinto, H.~Petersen, S.~Bass, M.~Keifer, S.~Zhong, C.~Healey, and R.~M.~T. Ii.
\newblock Comparative visualization of ensembles using ensemble surface slicing.
\newblock In {\em Visualization and {Data} {Analysis}}, vol. 8294, pp. 318--329. SPIE, Jan. 2012. \href{https://doi.org/10.1117/12.908288}
{doi: \textsf{%
10\hspace{.1pt}\discretionary{.}{%
}{.}\hspace{.4pt}1117\discretionary{/}{%
}{/}12\hspace{.1pt}\discretionary{.}{%
}{.}\hspace{.4pt}908288}}


\bibitem{arora_experimental_2017}
R.~Arora, R.~H. Kazi, F.~Anderson, T.~Grossman, K.~Singh, and G.~Fitzmaurice.
\newblock Experimental {Evaluation} of {Sketching} on {Surfaces} in {VR}.
\newblock In {\em Proceedings of the 2017 {CHI} {Conference} on {Human} {Factors} in {Computing} {Systems}}, pp. 5643--5654. ACM, Denver, May 2017. \href{https://doi.org/10.1145/3025453.3025474}
{doi: \textsf{%
10\hspace{.1pt}\discretionary{.}{%
}{.}\hspace{.4pt}1145\discretionary{/}{%
}{/}3025453\hspace{.1pt}\discretionary{.}{%
}{.}\hspace{.4pt}3025474}}


\bibitem{arora_symbiosissketch_2018}
R.~Arora, R.~H. Kazi, T.~Grossman, G.~Fitzmaurice, and K.~Singh.
\newblock {SymbiosisSketch}: {Combining} {2D} \& {3D} {Sketching} for {Designing} {Detailed} {3D} {Objects} in {Situ}.
\newblock In {\em Proceedings of the 2018 {CHI} {Conference} on {Human} {Factors} in {Computing} {Systems}}, pp. 1--15. ACM, Montreal, Apr. 2018. \href{https://doi.org/10.1145/3173574.3173759}
{doi: \textsf{%
10\hspace{.1pt}\discretionary{.}{%
}{.}\hspace{.4pt}1145\discretionary{/}{%
}{/}3173574\hspace{.1pt}\discretionary{.}{%
}{.}\hspace{.4pt}3173759}}


\bibitem{avila_haptic_1996}
R.~Avila and L.~Sobierajski.
\newblock A haptic interaction method for volume visualization.
\newblock In {\em Proceedings of {Seventh} {Annual} {IEEE} {Visualization}}, pp. 197--204, Oct. 1996. \href{https://doi.org/10.1109/VISUAL.1996.568108}
{doi: \textsf{%
10\hspace{.1pt}\discretionary{.}{%
}{.}\hspace{.4pt}1109\discretionary{/}{%
}{/}VISUAL\hspace{.1pt}\discretionary{.}{%
}{.}\hspace{.4pt}1996\hspace{.1pt}\discretionary{.}{%
}{.}\hspace{.4pt}568108}}


\bibitem{bae_ilovesketch_2008}
S.-H. Bae, R.~Balakrishnan, and K.~Singh.
\newblock {ILoveSketch}: as-natural-as-possible sketching system for creating 3d curve models.
\newblock In {\em Proceedings of the 21st annual {ACM} symposium on {User} interface software and technology}, pp. 151--160. ACM, Monterey, Oct. 2008. \href{https://doi.org/10.1145/1449715.1449740}
{doi: \textsf{%
10\hspace{.1pt}\discretionary{.}{%
}{.}\hspace{.4pt}1145\discretionary{/}{%
}{/}1449715\hspace{.1pt}\discretionary{.}{%
}{.}\hspace{.4pt}1449740}}


\bibitem{basdogan_haptic_2001}
C.~Basdogan and A.~Srinivasan.
\newblock Haptic {Rendering} in {Virtual} {Environments}.
\newblock Aug. 2001.

\bibitem{besancon_state_2021}
L.~Besançon, A.~Ynnerman, D.~F. Keefe, L.~Yu, and T.~Isenberg.
\newblock The {State} of the {Art} of {Spatial} {Interfaces} for {3D} {Visualization}.
\newblock {\em Computer Graphics Forum}, 40(1):293--326, 2021. \href{https://doi.org/10.1111/cgf.14189}
{doi: \textsf{%
10\hspace{.1pt}\discretionary{.}{%
}{.}\hspace{.4pt}1111\discretionary{/}{%
}{/}cgf\hspace{.1pt}\discretionary{.}{%
}{.}\hspace{.4pt}14189}}


\bibitem{blender}
{Blender Foundation}.
\newblock {Blender}.
\newblock \url{https://www.blender.org/download/releases/2-93/}, 2021.
\newblock Accessed: Mar 28, 2024.

\bibitem{bowman_virtual_2007}
D.~A. Bowman and R.~P. McMahan.
\newblock Virtual {Reality}: {How} {Much} {Immersion} {Is} {Enough}?
\newblock {\em Computer}, 40(7):36--43, July 2007. \href{https://doi.org/10.1109/MC.2007.257}
{doi: \textsf{%
10\hspace{.1pt}\discretionary{.}{%
}{.}\hspace{.4pt}1109\discretionary{/}{%
}{/}MC\hspace{.1pt}\discretionary{.}{%
}{.}\hspace{.4pt}2007\hspace{.1pt}\discretionary{.}{%
}{.}\hspace{.4pt}257}}


\bibitem{cazals_delaunay_2006}
F.~Cazals and J.~Giesen.
\newblock Delaunay {Triangulation} {Based} {Surface} {Reconstruction}.
\newblock In J.-D. Boissonnat and M.~Teillaud, eds., {\em Effective {Computational} {Geometry} for {Curves} and {Surfaces}}, pp. 231--276. Springer, Berlin, 2006. \href{https://doi.org/10.1007/978-3-540-33259-6_6}
{doi: \textsf{%
10\hspace{.1pt}\discretionary{.}{%
}{.}\hspace{.4pt}1007\discretionary{/}{%
}{/}978\discretionary{%
}{-}{-}3\discretionary{%
}{-}{-}540\discretionary{%
}{-}{-}33259\discretionary{%
}{-}{-}6\_6}}


\bibitem{chen_benchmark_2009}
X.~Chen, A.~Golovinskiy, and T.~Funkhouser.
\newblock A benchmark for {3D} mesh segmentation.
\newblock {\em ACM Transactions on Graphics}, 28(3):73:1--73:12, July 2009. \href{https://doi.org/10.1145/1531326.1531379}
{doi: \textsf{%
10\hspace{.1pt}\discretionary{.}{%
}{.}\hspace{.4pt}1145\discretionary{/}{%
}{/}1531326\hspace{.1pt}\discretionary{.}{%
}{.}\hspace{.4pt}1531379}}


\bibitem{choi_force_2005}
S.~Choi, L.~Walker, H.~Z. Tan, S.~Crittenden, and R.~Reifenberger.
\newblock Force constancy and its effect on haptic perception of virtual surfaces.
\newblock {\em ACM Transactions on Applied Perception}, 2(2):89--105, Apr. 2005. \href{https://doi.org/10.1145/1060581.1060584}
{doi: \textsf{%
10\hspace{.1pt}\discretionary{.}{%
}{.}\hspace{.4pt}1145\discretionary{/}{%
}{/}1060581\hspace{.1pt}\discretionary{.}{%
}{.}\hspace{.4pt}1060584}}


\bibitem{claussen_real_1992}
U.~Claussen.
\newblock Real {Time} {Phong} {Shading}.
\newblock In R.~L. Grimsdale and A.~Kaufman, eds., {\em Advances in {Computer} {Graphics} {Hardware} {V}: {Rendering}, {Ray} {Tracing} and {Visualization} {Systems}}, pp. 29--37. Springer, Berlin, 1992.

\bibitem{corenthy_volume_2015}
L.~Corenthy, M.~A. Otaduy, L.~Pastor, and M.~Garcia.
\newblock Volume {Haptics} with {Topology}-{Consistent} {Isosurfaces}.
\newblock {\em IEEE Transactions on Haptics}, 8(4):480--491, Oct. 2015. \href{https://doi.org/10.1109/TOH.2015.2466239}
{doi: \textsf{%
10\hspace{.1pt}\discretionary{.}{%
}{.}\hspace{.4pt}1109\discretionary{/}{%
}{/}TOH\hspace{.1pt}\discretionary{.}{%
}{.}\hspace{.4pt}2015\hspace{.1pt}\discretionary{.}{%
}{.}\hspace{.4pt}2466239}}


\bibitem{cumming_inference_2009}
G.~Cumming.
\newblock Inference by eye: {Reading} the overlap of independent confidence intervals.
\newblock {\em Statistics in Medicine}, 28(2):205--220, 2009. \href{https://doi.org/10.1002/sim.3471}
{doi: \textsf{%
10\hspace{.1pt}\discretionary{.}{%
}{.}\hspace{.4pt}1002\discretionary{/}{%
}{/}sim\hspace{.1pt}\discretionary{.}{%
}{.}\hspace{.4pt}3471}}


\bibitem{dragicevic_fair_2016}
P.~Dragicevic.
\newblock Fair {Statistical} {Communication} in {HCI}.
\newblock In J.~Robertson and M.~Kaptein, eds., {\em Modern {Statistical} {Methods} for {HCI}}, pp. 291--330. Springer International Publishing, Cham, 2016. \href{https://doi.org/10.1007/978-3-319-26633-6_13}
{doi: \textsf{%
10\hspace{.1pt}\discretionary{.}{%
}{.}\hspace{.4pt}1007\discretionary{/}{%
}{/}978\discretionary{%
}{-}{-}3\discretionary{%
}{-}{-}319\discretionary{%
}{-}{-}26633\discretionary{%
}{-}{-}6\_13}}


\bibitem{duriez_realistic_2006}
C.~Duriez, F.~Dubois, A.~Kheddar, and C.~Andriot.
\newblock Realistic haptic rendering of interacting deformable objects in virtual environments.
\newblock {\em IEEE Transactions on Visualization and Computer Graphics}, 12(1):36--47, Jan. 2006. \href{https://doi.org/10.1109/TVCG.2006.13}
{doi: \textsf{%
10\hspace{.1pt}\discretionary{.}{%
}{.}\hspace{.4pt}1109\discretionary{/}{%
}{/}TVCG\hspace{.1pt}\discretionary{.}{%
}{.}\hspace{.4pt}2006\hspace{.1pt}\discretionary{.}{%
}{.}\hspace{.4pt}13}}


\bibitem{edmunds_surface-based_2012}
M.~Edmunds, R.~S. Laramee, G.~Chen, N.~Max, E.~Zhang, and C.~Ware.
\newblock Surface-based flow visualization.
\newblock {\em Computers \& Graphics}, 36(8):974--990, Dec. 2012. \href{https://doi.org/10.1016/j.cag.2012.07.006}
{doi: \textsf{%
10\hspace{.1pt}\discretionary{.}{%
}{.}\hspace{.4pt}1016\discretionary{/}{%
}{/}j\hspace{.1pt}\discretionary{.}{%
}{.}\hspace{.4pt}cag\hspace{.1pt}\discretionary{.}{%
}{.}\hspace{.4pt}2012\hspace{.1pt}\discretionary{.}{%
}{.}\hspace{.4pt}07\hspace{.1pt}\discretionary{.}{%
}{.}\hspace{.4pt}006}}


\bibitem{elmqvist_taxonomy_2008}
N.~Elmqvist and P.~Tsigas.
\newblock A {Taxonomy} of {3D} {Occlusion} {Management} for {Visualization}.
\newblock {\em IEEE Transactions on Visualization and Computer Graphics}, 14(5):1095--1109, Sept. 2008. \href{https://doi.org/10.1109/TVCG.2008.59}
{doi: \textsf{%
10\hspace{.1pt}\discretionary{.}{%
}{.}\hspace{.4pt}1109\discretionary{/}{%
}{/}TVCG\hspace{.1pt}\discretionary{.}{%
}{.}\hspace{.4pt}2008\hspace{.1pt}\discretionary{.}{%
}{.}\hspace{.4pt}59}}


\bibitem{elsayed_vrsketchpen_2020}
H.~Elsayed, M.~D. Barrera~Machuca, C.~Schaarschmidt, K.~Marky, F.~Müller, J.~Riemann, A.~Matviienko, M.~Schmitz, M.~Weigel, and M.~Mühlhäuser.
\newblock {VRSketchPen}: {Unconstrained} {Haptic} {Assistance} for {Sketching} in {Virtual} {3D} {Environments}.
\newblock In {\em 26th {ACM} {Symposium} on {Virtual} {Reality} {Software} and {Technology}}, pp. 3:1--3:11. ACM, Canada, Nov. 2020. \href{https://doi.org/10.1145/3385956.3418953}
{doi: \textsf{%
10\hspace{.1pt}\discretionary{.}{%
}{.}\hspace{.4pt}1145\discretionary{/}{%
}{/}3385956\hspace{.1pt}\discretionary{.}{%
}{.}\hspace{.4pt}3418953}}


\bibitem{englund_touching_2018}
R.~Englund, K.~L. Palmerius, I.~Hotz, and A.~Ynnerman.
\newblock Touching {Data}: {Enhancing} {Visual} {Exploration} of {Flow} {Data} with {Haptics}.
\newblock {\em Computing in Science \& Engineering}, 20(3):89--100, May 2018. \href{https://doi.org/10.1109/MCSE.2018.03221931}
{doi: \textsf{%
10\hspace{.1pt}\discretionary{.}{%
}{.}\hspace{.4pt}1109\discretionary{/}{%
}{/}MCSE\hspace{.1pt}\discretionary{.}{%
}{.}\hspace{.4pt}2018\hspace{.1pt}\discretionary{.}{%
}{.}\hspace{.4pt}03221931}}


\bibitem{faeth_combining_2008}
A.~Faeth, M.~Oren, and C.~Harding.
\newblock Combining 3-{D} geovisualization with force feedback driven user interaction.
\newblock In {\em Proceedings of the 16th {ACM} {SIGSPATIAL} international conference on {Advances} in geographic information systems}, pp. 25:1--25:9. ACM, Irvine, Nov. 2008. \href{https://doi.org/10.1145/1463434.1463466}
{doi: \textsf{%
10\hspace{.1pt}\discretionary{.}{%
}{.}\hspace{.4pt}1145\discretionary{/}{%
}{/}1463434\hspace{.1pt}\discretionary{.}{%
}{.}\hspace{.4pt}1463466}}


\bibitem{fonnet_survey_2021}
A.~Fonnet and Y.~Prie.
\newblock Survey of {Immersive} {Analytics}.
\newblock {\em IEEE Transactions on Visualization and Computer Graphics}, 27(3):2101--2122, Mar. 2021. \href{https://doi.org/10.1109/TVCG.2019.2929033}
{doi: \textsf{%
10\hspace{.1pt}\discretionary{.}{%
}{.}\hspace{.4pt}1109\discretionary{/}{%
}{/}TVCG\hspace{.1pt}\discretionary{.}{%
}{.}\hspace{.4pt}2019\hspace{.1pt}\discretionary{.}{%
}{.}\hspace{.4pt}2929033}}


\bibitem{gibson_perception_1950}
J.~J. Gibson.
\newblock The {Perception} of {Visual} {Surfaces}.
\newblock {\em The American Journal of Psychology}, 63(3):367--384, 1950. \href{https://doi.org/10.2307/1418003}
{doi: \textsf{%
10\hspace{.1pt}\discretionary{.}{%
}{.}\hspace{.4pt}2307\discretionary{/}{%
}{/}1418003}}


\bibitem{grossman_creating_2002}
T.~Grossman, R.~Balakrishnan, G.~Kurtenbach, G.~Fitzmaurice, A.~Khan, and B.~Buxton.
\newblock Creating principal {3D} curves with digital tape drawing.
\newblock In {\em Proceedings of the {SIGCHI} {Conference} on {Human} {Factors} in {Computing} {Systems}}, pp. 121--128. ACM, Minneapolis, Apr. 2002. \href{https://doi.org/10.1145/503376.503398}
{doi: \textsf{%
10\hspace{.1pt}\discretionary{.}{%
}{.}\hspace{.4pt}1145\discretionary{/}{%
}{/}503376\hspace{.1pt}\discretionary{.}{%
}{.}\hspace{.4pt}503398}}


\bibitem{hoffmann_thermalpen_2023}
P.~P. Hoffmann, H.~Elsayed, M.~Mühlhäuser, R.~R. Wehbe, and M.~D. Barrera~Machuca.
\newblock {ThermalPen}: {Adding} {Thermal} {Haptic} {Feedback} to {3D} {Sketching}.
\newblock In {\em Extended {Abstracts} of the 2023 {CHI} {Conference} on {Human} {Factors} in {Computing} {Systems}}, pp. 474:1--474:4. ACM, Hamburg, Apr. 2023. \href{https://doi.org/10.1145/3544549.3583901}
{doi: \textsf{%
10\hspace{.1pt}\discretionary{.}{%
}{.}\hspace{.4pt}1145\discretionary{/}{%
}{/}3544549\hspace{.1pt}\discretionary{.}{%
}{.}\hspace{.4pt}3583901}}


\bibitem{ikits_constraint-based_2003}
M.~Ikits, J.~Brederson, C.~Hansen, and C.~Johnson.
\newblock A constraint-based technique for haptic volume exploration.
\newblock In {\em {IEEE} {Visualization}, 2003. {VIS} 2003}, pp. 263--269. IEEE, Seattle, 2003. \href{https://doi.org/10.1109/VISUAL.2003.1250381}
{doi: \textsf{%
10\hspace{.1pt}\discretionary{.}{%
}{.}\hspace{.4pt}1109\discretionary{/}{%
}{/}VISUAL\hspace{.1pt}\discretionary{.}{%
}{.}\hspace{.4pt}2003\hspace{.1pt}\discretionary{.}{%
}{.}\hspace{.4pt}1250381}}


\bibitem{itkowitz_openhaptics_2005}
B.~Itkowitz, J.~Handley, and {Weihang Zhu}.
\newblock The {OpenHaptics}™ {Toolkit}: {A} {Library} for {Adding} {3D} {Touch}™ {Navigation} and {Haptics} to {Graphics} {Applications}.
\newblock In {\em First {Joint} {Eurohaptics} {Conference} and {Symposium} on {Haptic} {Interfaces} for {Virtual} {Environment} and {Teleoperator} {Systems}}, pp. 590--591. IEEE, Pisa, 2005. \href{https://doi.org/10.1109/WHC.2005.133}
{doi: \textsf{%
10\hspace{.1pt}\discretionary{.}{%
}{.}\hspace{.4pt}1109\discretionary{/}{%
}{/}WHC\hspace{.1pt}\discretionary{.}{%
}{.}\hspace{.4pt}2005\hspace{.1pt}\discretionary{.}{%
}{.}\hspace{.4pt}133}}


\bibitem{jackson_force_2012}
B.~Jackson, D.~Coffey, and D.~F. Keefe.
\newblock Force {Brushes}: {Progressive} {Data}-{Driven} {Haptic} {Selection} and {Filtering} for {Multi}-{Variate} {Flow} {Visualizations}.
\newblock {\em EuroVis - Short Papers}, pp. 7--11, 2012. \href{https://doi.org//10.2312/PE/EuroVisShort/EuroVisShort2012/007-011}
{doi: \textsf{%
\discretionary{/}{%
}{/}10\hspace{.1pt}\discretionary{.}{%
}{.}\hspace{.4pt}2312\discretionary{/}{%
}{/}PE\discretionary{/}{%
}{/}EuroVisShort\discretionary{/}{%
}{/}EuroVisShort2012\discretionary{/}{%
}{/}007\discretionary{%
}{-}{-}011}}


\bibitem{jackson_lift-off_2016}
B.~Jackson and D.~F. Keefe.
\newblock Lift-{Off}: {Using} {Reference} {Imagery} and {Freehand} {Sketching} to {Create} {3D} {Models} in {VR}.
\newblock {\em IEEE Transactions on Visualization and Computer Graphics}, 22(4):1442--1451, Apr. 2016. \href{https://doi.org/10.1109/TVCG.2016.2518099}
{doi: \textsf{%
10\hspace{.1pt}\discretionary{.}{%
}{.}\hspace{.4pt}1109\discretionary{/}{%
}{/}TVCG\hspace{.1pt}\discretionary{.}{%
}{.}\hspace{.4pt}2016\hspace{.1pt}\discretionary{.}{%
}{.}\hspace{.4pt}2518099}}


\bibitem{keefe_drawing_2007}
D.~Keefe, R.~Zeleznik, and D.~Laidlaw.
\newblock Drawing on {Air}: {Input} {Techniques} for {Controlled} {3D} {Line} {Illustration}.
\newblock {\em IEEE Transactions on Visualization and Computer Graphics}, 13(5):1067--1081, Sept. 2007. \href{https://doi.org/10.1109/TVCG.2007.1060}
{doi: \textsf{%
10\hspace{.1pt}\discretionary{.}{%
}{.}\hspace{.4pt}1109\discretionary{/}{%
}{/}TVCG\hspace{.1pt}\discretionary{.}{%
}{.}\hspace{.4pt}2007\hspace{.1pt}\discretionary{.}{%
}{.}\hspace{.4pt}1060}}


\bibitem{kim_agile_2018}
Y.~Kim, S.-G. An, J.~H. Lee, and S.-H. Bae.
\newblock Agile {3D} {Sketching} with {Air} {Scaffolding}.
\newblock In {\em Proceedings of the 2018 {CHI} {Conference} on {Human} {Factors} in {Computing} {Systems}}, pp. 238:1--238:12. ACM, Montreal, Apr. 2018. \href{https://doi.org/10.1145/3173574.3173812}
{doi: \textsf{%
10\hspace{.1pt}\discretionary{.}{%
}{.}\hspace{.4pt}1145\discretionary{/}{%
}{/}3173574\hspace{.1pt}\discretionary{.}{%
}{.}\hspace{.4pt}3173812}}


\bibitem{kinnear_procedural_2010}
K.~Kinnear and C.~S. Kaplan.
\newblock Procedural generation of surface detail for science ﬁction spaceships.
\newblock {\em Computational Aesthetics in Graphics, Visualization, and Imaging}, 2010. \href{https://doi.org//10.2312/COMPAESTH/COMPAESTH10/083-090}
{doi: \textsf{%
\discretionary{/}{%
}{/}10\hspace{.1pt}\discretionary{.}{%
}{.}\hspace{.4pt}2312\discretionary{/}{%
}{/}COMPAESTH\discretionary{/}{%
}{/}COMPAESTH10\discretionary{/}{%
}{/}083\discretionary{%
}{-}{-}090}}


\bibitem{komerska_haptic_2004}
R.~Komerska and C.~Ware.
\newblock Haptic state surface interactions.
\newblock {\em IEEE Computer Graphics and Applications}, 24(6):52--59, Nov. 2004. \href{https://doi.org/10.1109/MCG.2004.53}
{doi: \textsf{%
10\hspace{.1pt}\discretionary{.}{%
}{.}\hspace{.4pt}1109\discretionary{/}{%
}{/}MCG\hspace{.1pt}\discretionary{.}{%
}{.}\hspace{.4pt}2004\hspace{.1pt}\discretionary{.}{%
}{.}\hspace{.4pt}53}}


\bibitem{laehyun_kim_haptic_2004}
{Laehyun Kim}, G.~Sukhatme, and M.~Desbrun.
\newblock Haptic rendering-beyond visual computing - {A} haptic-rendering technique based on hybrid surface representation.
\newblock {\em IEEE Computer Graphics and Applications}, 24(2):66--75, Mar. 2004. \href{https://doi.org/10.1109/MCG.2004.1274064}
{doi: \textsf{%
10\hspace{.1pt}\discretionary{.}{%
}{.}\hspace{.4pt}1109\discretionary{/}{%
}{/}MCG\hspace{.1pt}\discretionary{.}{%
}{.}\hspace{.4pt}2004\hspace{.1pt}\discretionary{.}{%
}{.}\hspace{.4pt}1274064}}


\bibitem{laha_effects_2014}
B.~Laha, D.~A. Bowman, and J.~J. Socha.
\newblock Effects of {VR} {System} {Fidelity} on {Analyzing} {Isosurface} {Visualization} of {Volume} {Datasets}.
\newblock {\em IEEE Transactions on Visualization and Computer Graphics}, 20(4):513--522, Apr. 2014. \href{https://doi.org/10.1109/TVCG.2014.20}
{doi: \textsf{%
10\hspace{.1pt}\discretionary{.}{%
}{.}\hspace{.4pt}1109\discretionary{/}{%
}{/}TVCG\hspace{.1pt}\discretionary{.}{%
}{.}\hspace{.4pt}2014\hspace{.1pt}\discretionary{.}{%
}{.}\hspace{.4pt}20}}


\bibitem{laha_effects_2012}
B.~Laha, K.~Sensharma, J.~D. Schiffbauer, and D.~A. Bowman.
\newblock Effects of {Immersion} on {Visual} {Analysis} of {Volume} {Data}.
\newblock {\em IEEE Transactions on Visualization and Computer Graphics}, 18(4):597--606, Apr. 2012. \href{https://doi.org/10.1109/TVCG.2012.42}
{doi: \textsf{%
10\hspace{.1pt}\discretionary{.}{%
}{.}\hspace{.4pt}1109\discretionary{/}{%
}{/}TVCG\hspace{.1pt}\discretionary{.}{%
}{.}\hspace{.4pt}2012\hspace{.1pt}\discretionary{.}{%
}{.}\hspace{.4pt}42}}


\bibitem{lawonn_improving_2017}
K.~Lawonn, M.~Luz, and C.~Hansen.
\newblock Improving spatial perception of vascular models using supporting anchors and illustrative visualization.
\newblock {\em Computers \& Graphics}, 63:37--49, Apr. 2017. \href{https://doi.org/10.1016/j.cag.2017.02.002}
{doi: \textsf{%
10\hspace{.1pt}\discretionary{.}{%
}{.}\hspace{.4pt}1016\discretionary{/}{%
}{/}j\hspace{.1pt}\discretionary{.}{%
}{.}\hspace{.4pt}cag\hspace{.1pt}\discretionary{.}{%
}{.}\hspace{.4pt}2017\hspace{.1pt}\discretionary{.}{%
}{.}\hspace{.4pt}02\hspace{.1pt}\discretionary{.}{%
}{.}\hspace{.4pt}002}}


\bibitem{lawonn_survey_2018}
K.~Lawonn, I.~Viola, B.~Preim, and T.~Isenberg.
\newblock A {Survey} of {Surface}-{Based} {Illustrative} {Rendering} for {Visualization}.
\newblock {\em Computer Graphics Forum}, 37(6):205--234, 2018. \href{https://doi.org/10.1111/cgf.13322}
{doi: \textsf{%
10\hspace{.1pt}\discretionary{.}{%
}{.}\hspace{.4pt}1111\discretionary{/}{%
}{/}cgf\hspace{.1pt}\discretionary{.}{%
}{.}\hspace{.4pt}13322}}


\bibitem{li_visibility_2017}
N.~Li, W.~Willett, E.~Sharlin, and M.~C. Sousa.
\newblock Visibility perception and dynamic viewsheds for topographic maps and models.
\newblock In {\em Proceedings of the 5th {Symposium} on {Spatial} {User} {Interaction}}, {SUI} '17, pp. 39--47. ACM, New York, Oct. 2017. \href{https://doi.org/10.1145/3131277.3132178}
{doi: \textsf{%
10\hspace{.1pt}\discretionary{.}{%
}{.}\hspace{.4pt}1145\discretionary{/}{%
}{/}3131277\hspace{.1pt}\discretionary{.}{%
}{.}\hspace{.4pt}3132178}}


\bibitem{lin_haptic_2008}
M.~C. Lin and M.~Otaduy.
\newblock {\em Haptic {Rendering}: {Foundations}, {Algorithms}, and {Applications}}.
\newblock CRC Press, July 2008.

\bibitem{lorensen_marching_1987}
W.~E. Lorensen and H.~E. Cline.
\newblock Marching cubes: {A} high resolution {3D} surface construction algorithm.
\newblock In {\em Proceedings of the 14th annual conference on {Computer} graphics and interactive techniques}, {SIGGRAPH} '87, pp. 163--169. ACM, New York, Aug. 1987. \href{https://doi.org/10.1145/37401.37422}
{doi: \textsf{%
10\hspace{.1pt}\discretionary{.}{%
}{.}\hspace{.4pt}1145\discretionary{/}{%
}{/}37401\hspace{.1pt}\discretionary{.}{%
}{.}\hspace{.4pt}37422}}


\bibitem{lorenz_perceived_2023}
M.~Lorenz, A.~Hoffmann, M.~Kaluschke, T.~Ziadeh, N.~Pillen, M.~Kusserow, J.~Perret, S.~Knopp, A.~Dettmann, P.~Klimant, G.~Zachmann, and A.~C. Bullinger.
\newblock Perceived realism of haptic rendering methods for bimanual high force tasks: original and replication study.
\newblock {\em Scientific Reports}, 13(1):11230, July 2023. \href{https://doi.org/10.1038/s41598-023-38201-x}
{doi: \textsf{%
10\hspace{.1pt}\discretionary{.}{%
}{.}\hspace{.4pt}1038\discretionary{/}{%
}{/}s41598\discretionary{%
}{-}{-}023\discretionary{%
}{-}{-}38201\discretionary{%
}{-}{-}x}}


\bibitem{machuca_multiplanes_2018}
M.~D.~B. Machuca, P.~Asente, W.~Stuerzlinger, J.~Lu, and B.~Kim.
\newblock Multiplanes: {Assisted} {Freehand} {VR} {Sketching}.
\newblock In {\em Proceedings of the {Symposium} on {Spatial} {User} {Interaction}}, pp. 36--47. ACM, Berlin, Oct. 2018. \href{https://doi.org/10.1145/3267782.3267786}
{doi: \textsf{%
10\hspace{.1pt}\discretionary{.}{%
}{.}\hspace{.4pt}1145\discretionary{/}{%
}{/}3267782\hspace{.1pt}\discretionary{.}{%
}{.}\hspace{.4pt}3267786}}


\bibitem{mackenzie_chapter_2013}
I.~S. MacKenzie.
\newblock Chapter 5 - {Designing} {HCI} {Experiments}.
\newblock In I.~S. MacKenzie, ed., {\em Human-computer {Interaction}}, pp. 157--189. MK, Boston, Jan. 2013.

\bibitem{meuschke_evalviz_2019}
M.~Meuschke, N.~N. Smit, N.~Lichtenberg, B.~Preim, and K.~Lawonn.
\newblock {EvalViz} – {Surface} visualization evaluation wizard for depth and shape perception tasks.
\newblock {\em Computers \& Graphics}, 82:250--263, Aug. 2019. \href{https://doi.org/10.1016/j.cag.2019.05.022}
{doi: \textsf{%
10\hspace{.1pt}\discretionary{.}{%
}{.}\hspace{.4pt}1016\discretionary{/}{%
}{/}j\hspace{.1pt}\discretionary{.}{%
}{.}\hspace{.4pt}cag\hspace{.1pt}\discretionary{.}{%
}{.}\hspace{.4pt}2019\hspace{.1pt}\discretionary{.}{%
}{.}\hspace{.4pt}05\hspace{.1pt}\discretionary{.}{%
}{.}\hspace{.4pt}022}}


\bibitem{mills_geospatial_2008}
J.~W. Mills.
\newblock Geospatial {Analysis}: {A} {Comprehensive} {Guide} to {Principles}, {Techniques}, and {Software} {Tools}, {Second} {Edition} - by {Michael} {J}. de {Smith}, {Michael} {F}. {Goodchild}, and {Paul} {A}. {Longley}.
\newblock {\em Transactions in GIS}, 12(5):645--647, 2008. \href{https://doi.org/10.1111/j.1467-9671.2008.01122.x}
{doi: \textsf{%
10\hspace{.1pt}\discretionary{.}{%
}{.}\hspace{.4pt}1111\discretionary{/}{%
}{/}j\hspace{.1pt}\discretionary{.}{%
}{.}\hspace{.4pt}1467\discretionary{%
}{-}{-}9671\hspace{.1pt}\discretionary{.}{%
}{.}\hspace{.4pt}2008\hspace{.1pt}\discretionary{.}{%
}{.}\hspace{.4pt}01122\hspace{.1pt}\discretionary{.}{%
}{.}\hspace{.4pt}x}}


\bibitem{mittring_finding_2007}
M.~Mittring.
\newblock Finding next gen: {CryEngine} 2.
\newblock In {\em {ACM} {SIGGRAPH} 2007 courses}, pp. 97--121. ACM, San Diego, Aug. 2007. \href{https://doi.org/10.1145/1281500.1281671}
{doi: \textsf{%
10\hspace{.1pt}\discretionary{.}{%
}{.}\hspace{.4pt}1145\discretionary{/}{%
}{/}1281500\hspace{.1pt}\discretionary{.}{%
}{.}\hspace{.4pt}1281671}}


\bibitem{mohanty_investigating_2020}
R.~R. Mohanty, R.~M. Castillo, E.~D. Ragan, and V.~R. Krishnamurthy.
\newblock Investigating {Force}-{Feedback} in {Mid}-{Air} {Sketching} of {Multi}-{Planar} {Three}-{Dimensional} {Curve}-{Soups}.
\newblock {\em Journal of Computing and Information Science in Engineering}, 20(1):011010, Feb. 2020. \href{https://doi.org/10.1115/1.4045142}
{doi: \textsf{%
10\hspace{.1pt}\discretionary{.}{%
}{.}\hspace{.4pt}1115\discretionary{/}{%
}{/}1\hspace{.1pt}\discretionary{.}{%
}{.}\hspace{.4pt}4045142}}


\bibitem{moore_digital_1991}
I.~D. Moore, R.~B. Grayson, and A.~R. Ladson.
\newblock Digital terrain modelling: {A} review of hydrological, geomorphological, and biological applications.
\newblock {\em Hydrological Processes}, 5(1):3--30, 1991. \href{https://doi.org/10.1002/hyp.3360050103}
{doi: \textsf{%
10\hspace{.1pt}\discretionary{.}{%
}{.}\hspace{.4pt}1002\discretionary{/}{%
}{/}hyp\hspace{.1pt}\discretionary{.}{%
}{.}\hspace{.4pt}3360050103}}


\bibitem{nam_worlds--wedges_2019}
J.~W. Nam, K.~McCullough, J.~Tveite, M.~M. Espinosa, C.~H. Perry, B.~T. Wilson, and D.~F. Keefe.
\newblock Worlds-in-{Wedges}: {Combining} {Worlds}-in-{Miniature} and {Portals} to {Support} {Comparative} {Immersive} {Visualization} of {Forestry} {Data}.
\newblock In {\em 2019 {IEEE} {Conference} on {Virtual} {Reality} and {3D} {User} {Interfaces} ({VR})}, pp. 747--755. IEEE, Osaka, Mar. 2019. \href{https://doi.org/10.1109/VR.2019.8797871}
{doi: \textsf{%
10\hspace{.1pt}\discretionary{.}{%
}{.}\hspace{.4pt}1109\discretionary{/}{%
}{/}VR\hspace{.1pt}\discretionary{.}{%
}{.}\hspace{.4pt}2019\hspace{.1pt}\discretionary{.}{%
}{.}\hspace{.4pt}8797871}}


\bibitem{ohnishi_virtual_2012}
T.~Ohnishi, N.~Katzakis, K.~Kiyokawa, and H.~Takemura.
\newblock Virtual interaction surface: {Decoupling} of interaction and view dimensions for flexible indirect {3D} interaction.
\newblock In {\em 2012 {IEEE} {Symposium} on {3D} {User} {Interfaces} ({3DUI})}, pp. 113--116, Mar. 2012. \href{https://doi.org/10.1109/3DUI.2012.6184194}
{doi: \textsf{%
10\hspace{.1pt}\discretionary{.}{%
}{.}\hspace{.4pt}1109\discretionary{/}{%
}{/}3DUI\hspace{.1pt}\discretionary{.}{%
}{.}\hspace{.4pt}2012\hspace{.1pt}\discretionary{.}{%
}{.}\hspace{.4pt}6184194}}


\bibitem{paneels_review_2010}
S.~Paneels and J.~C. Roberts.
\newblock Review of {Designs} for {Haptic} {Data} {Visualization}.
\newblock {\em IEEE Transactions on Haptics}, 3(2):119--137, Apr. 2010. \href{https://doi.org/10.1109/TOH.2009.44}
{doi: \textsf{%
10\hspace{.1pt}\discretionary{.}{%
}{.}\hspace{.4pt}1109\discretionary{/}{%
}{/}TOH\hspace{.1pt}\discretionary{.}{%
}{.}\hspace{.4pt}2009\hspace{.1pt}\discretionary{.}{%
}{.}\hspace{.4pt}44}}


\bibitem{preim_survey_2016}
B.~Preim, A.~Baer, D.~Cunningham, T.~Isenberg, and T.~Ropinski.
\newblock A {Survey} of {Perceptually} {Motivated} {3D} {Visualization} of {Medical} {Image} {Data}.
\newblock {\em Computer Graphics Forum}, 35(3):501--525, 2016. \href{https://doi.org/10.1111/cgf.12927}
{doi: \textsf{%
10\hspace{.1pt}\discretionary{.}{%
}{.}\hspace{.4pt}1111\discretionary{/}{%
}{/}cgf\hspace{.1pt}\discretionary{.}{%
}{.}\hspace{.4pt}12927}}


\bibitem{prouzeau_scaptics_2019}
A.~Prouzeau, M.~Cordeil, C.~Robin, B.~Ens, B.~H. Thomas, and T.~Dwyer.
\newblock Scaptics and {Highlight}-{Planes}: {Immersive} {Interaction} {Techniques} for {Finding} {Occluded} {Features} in {3D} {Scatterplots}.
\newblock In {\em Proceedings of the 2019 {CHI} {Conference} on {Human} {Factors} in {Computing} {Systems}}, pp. 325:1--325:12. ACM, Glasgow, May 2019. \href{https://doi.org/10.1145/3290605.3300555}
{doi: \textsf{%
10\hspace{.1pt}\discretionary{.}{%
}{.}\hspace{.4pt}1145\discretionary{/}{%
}{/}3290605\hspace{.1pt}\discretionary{.}{%
}{.}\hspace{.4pt}3300555}}


\bibitem{perlin}
{Raouf}.
\newblock {Perlin Noise: A Procedural Generation Algorithm}.
\newblock \url{https://rtouti.github.io/graphics/perlin-noise-algorithm}.
\newblock Accessed: Jun 30, 2024.

\bibitem{reiner_seeing_2008}
M.~Reiner.
\newblock Seeing {Through} {Touch}: {The} {Role} of {Haptic} {Information} in {Visualization}.
\newblock In J.~K. Gilbert, M.~Reiner, and M.~Nakhleh, eds., {\em Visualization: {Theory} and {Practice} in {Science} {Education}}, pp. 73--84. Springer Netherlands, Dordrecht, 2008. \href{https://doi.org/10.1007/978-1-4020-5267-5_4}
{doi: \textsf{%
10\hspace{.1pt}\discretionary{.}{%
}{.}\hspace{.4pt}1007\discretionary{/}{%
}{/}978\discretionary{%
}{-}{-}1\discretionary{%
}{-}{-}4020\discretionary{%
}{-}{-}5267\discretionary{%
}{-}{-}5\_4}}


\bibitem{rocha_decal-maps_2017}
A.~Rocha, U.~Alim, J.~D. Silva, and M.~C. Sousa.
\newblock Decal-{Maps}: {Real}-{Time} {Layering} of {Decals} on {Surfaces} for {Multivariate} {Visualization}.
\newblock {\em IEEE transactions on visualization and computer graphics}, 23(1):821--830, Jan. 2017. \href{https://doi.org/10.1109/TVCG.2016.2598866}
{doi: \textsf{%
10\hspace{.1pt}\discretionary{.}{%
}{.}\hspace{.4pt}1109\discretionary{/}{%
}{/}TVCG\hspace{.1pt}\discretionary{.}{%
}{.}\hspace{.4pt}2016\hspace{.1pt}\discretionary{.}{%
}{.}\hspace{.4pt}2598866}}


\bibitem{rocha_decal-lenses_2019}
A.~Rocha, J.~D. Silva, U.~R. Alim, S.~Carpendale, and M.~C. Sousa.
\newblock Decal-{Lenses}: {Interactive} {Lenses} on {Surfaces} for {Multivariate} {Visualization}.
\newblock {\em IEEE Transactions on Visualization and Computer Graphics}, 25(8):2568--2582, Aug. 2019. \href{https://doi.org/10.1109/TVCG.2018.2850781}
{doi: \textsf{%
10\hspace{.1pt}\discretionary{.}{%
}{.}\hspace{.4pt}1109\discretionary{/}{%
}{/}TVCG\hspace{.1pt}\discretionary{.}{%
}{.}\hspace{.4pt}2018\hspace{.1pt}\discretionary{.}{%
}{.}\hspace{.4pt}2850781}}


\bibitem{ropinski_visually_2006}
T.~Ropinski, F.~Steinicke, and K.~Hinrichs.
\newblock Visually {Supporting} {Depth} {Perception} in {Angiography} {Imaging}.
\newblock In A.~Butz, B.~Fisher, A.~Krüger, and P.~Olivier, eds., {\em Smart {Graphics}}, pp. 93--104. Springer, Berlin, 2006. \href{https://doi.org/10.1007/11795018_9}
{doi: \textsf{%
10\hspace{.1pt}\discretionary{.}{%
}{.}\hspace{.4pt}1007\discretionary{/}{%
}{/}11795018\_9}}


\bibitem{ruspini_haptic_1997}
D.~C. Ruspini, K.~Kolarov, and O.~Khatib.
\newblock The haptic display of complex graphical environments.
\newblock In {\em Proceedings of the 24th annual conference on {Computer} graphics and interactive techniques - {SIGGRAPH} '97}, pp. 345--352. ACM Press, 1997. \href{https://doi.org/10.1145/258734.258878}
{doi: \textsf{%
10\hspace{.1pt}\discretionary{.}{%
}{.}\hspace{.4pt}1145\discretionary{/}{%
}{/}258734\hspace{.1pt}\discretionary{.}{%
}{.}\hspace{.4pt}258878}}


\bibitem{sharples_virtual_2008}
S.~Sharples, S.~Cobb, A.~Moody, and J.~R. Wilson.
\newblock Virtual reality induced symptoms and effects ({VRISE}): {Comparison} of head mounted display ({HMD}), desktop and projection display systems.
\newblock {\em Displays}, 29(2):58--69, Mar. 2008. \href{https://doi.org/10.1016/j.displa.2007.09.005}
{doi: \textsf{%
10\hspace{.1pt}\discretionary{.}{%
}{.}\hspace{.4pt}1016\discretionary{/}{%
}{/}j\hspace{.1pt}\discretionary{.}{%
}{.}\hspace{.4pt}displa\hspace{.1pt}\discretionary{.}{%
}{.}\hspace{.4pt}2007\hspace{.1pt}\discretionary{.}{%
}{.}\hspace{.4pt}09\hspace{.1pt}\discretionary{.}{%
}{.}\hspace{.4pt}005}}


\bibitem{song_wysiwyf_2011}
P.~Song, W.~B. Goh, C.-W. Fu, Q.~Meng, and P.-A. Heng.
\newblock {WYSIWYF}: exploring and annotating volume data with a tangible handheld device.
\newblock In {\em Proceedings of the {SIGCHI} {Conference} on {Human} {Factors} in {Computing} {Systems}}, {CHI} '11, pp. 1333--1342. ACM, New York, May 2011. \href{https://doi.org/10.1145/1978942.1979140}
{doi: \textsf{%
10\hspace{.1pt}\discretionary{.}{%
}{.}\hspace{.4pt}1145\discretionary{/}{%
}{/}1978942\hspace{.1pt}\discretionary{.}{%
}{.}\hspace{.4pt}1979140}}


\bibitem{strohmeier_generating_2017}
P.~Strohmeier and K.~Hornbæk.
\newblock Generating {Haptic} {Textures} with a {Vibrotactile} {Actuator}.
\newblock In {\em Proceedings of the 2017 {CHI} {Conference} on {Human} {Factors} in {Computing} {Systems}}, {CHI} '17, pp. 4994--5005. ACM, New York, May 2017. \href{https://doi.org/10.1145/3025453.3025812}
{doi: \textsf{%
10\hspace{.1pt}\discretionary{.}{%
}{.}\hspace{.4pt}1145\discretionary{/}{%
}{/}3025453\hspace{.1pt}\discretionary{.}{%
}{.}\hspace{.4pt}3025812}}


\bibitem{webster_reconstruction_2015}
K.~Sunderland, B.~Woo, C.~Pinter, and G.~Fichtinger.
\newblock Reconstruction of surfaces from planar contours through contour interpolation.
\newblock p. 94151R. Orlando, Mar. 2015. \href{https://doi.org/10.1117/12.2081436}
{doi: \textsf{%
10\hspace{.1pt}\discretionary{.}{%
}{.}\hspace{.4pt}1117\discretionary{/}{%
}{/}12\hspace{.1pt}\discretionary{.}{%
}{.}\hspace{.4pt}2081436}}


\bibitem{tietjen_combining_2005}
C.~Tietjen, T.~Isenberg, and B.~Preim.
\newblock {\em Combining {Silhouettes}, {Surface}, and {Volume} {Rendering} for {Surgery} {Education} and {Planning}}.
\newblock The Eurographics Association, 2005. \href{https://doi.org/10.2312/VisSym/EuroVis05/303-310}
{doi: \textsf{%
10\hspace{.1pt}\discretionary{.}{%
}{.}\hspace{.4pt}2312\discretionary{/}{%
}{/}VisSym\discretionary{/}{%
}{/}EuroVis05\discretionary{/}{%
}{/}303\discretionary{%
}{-}{-}310}}


\bibitem{top_spotlight_2011}
A.~Top, G.~Hamarneh, and R.~Abugharbieh.
\newblock Spotlight: {Automated} {Confidence}-{Based} {User} {Guidance} for {Increasing} {Efficiency} in {Interactive} {3D} {Image} {Segmentation}.
\newblock In B.~Menze, G.~Langs, Z.~Tu, and A.~Criminisi, eds., {\em Medical {Computer} {Vision}. {Recognition} {Techniques} and {Applications} in {Medical} {Imaging}}, pp. 204--213. Springer, Berlin, 2011. \href{https://doi.org/10.1007/978-3-642-18421-5_20}
{doi: \textsf{%
10\hspace{.1pt}\discretionary{.}{%
}{.}\hspace{.4pt}1007\discretionary{/}{%
}{/}978\discretionary{%
}{-}{-}3\discretionary{%
}{-}{-}642\discretionary{%
}{-}{-}18421\discretionary{%
}{-}{-}5\_20}}


\bibitem{usher_virtual_2018}
W.~Usher, P.~Klacansky, F.~Federer, P.-T. Bremer, A.~Knoll, J.~Yarch, A.~Angelucci, and V.~Pascucci.
\newblock A {Virtual} {Reality} {Visualization} {Tool} for {Neuron} {Tracing}.
\newblock {\em IEEE Transactions on Visualization and Computer Graphics}, 24(1):994--1003, Jan. 2018. \href{https://doi.org/10.1109/TVCG.2017.2744079}
{doi: \textsf{%
10\hspace{.1pt}\discretionary{.}{%
}{.}\hspace{.4pt}1109\discretionary{/}{%
}{/}TVCG\hspace{.1pt}\discretionary{.}{%
}{.}\hspace{.4pt}2017\hspace{.1pt}\discretionary{.}{%
}{.}\hspace{.4pt}2744079}}


\bibitem{wang_haptic_2014}
D.~Wang, J.~Xiao, and Y.~Zhang.
\newblock {\em Haptic {Rendering} for {Simulation} of {Fine} {Manipulation}}.
\newblock Springer, Berlin, 2014. \href{https://doi.org/10.1007/978-3-662-44949-3}
{doi: \textsf{%
10\hspace{.1pt}\discretionary{.}{%
}{.}\hspace{.4pt}1007\discretionary{/}{%
}{/}978\discretionary{%
}{-}{-}3\discretionary{%
}{-}{-}662\discretionary{%
}{-}{-}44949\discretionary{%
}{-}{-}3}}


\bibitem{wang_understanding_2022}
X.~Wang, L.~Besançon, M.~Ammi, and T.~Isenberg.
\newblock Understanding differences between combinations of {2D} and {3D} input and output devices for {3D} data visualization.
\newblock {\em International Journal of Human-Computer Studies}, 163:102820, July 2022. \href{https://doi.org/10.1016/j.ijhcs.2022.102820}
{doi: \textsf{%
10\hspace{.1pt}\discretionary{.}{%
}{.}\hspace{.4pt}1016\discretionary{/}{%
}{/}j\hspace{.1pt}\discretionary{.}{%
}{.}\hspace{.4pt}ijhcs\hspace{.1pt}\discretionary{.}{%
}{.}\hspace{.4pt}2022\hspace{.1pt}\discretionary{.}{%
}{.}\hspace{.4pt}102820}}


\bibitem{ware_view_nodate}
C.~Ware and G.~Sweet.
\newblock View {Direction}, {Surface} {Orientation} and {Texture} {Orientation} for {Perception} of {Surface} {Shape}.
\newblock {\em Proceedings of Graphics Interface 2004}, pp. 97--106. \href{https://doi.org/10.1145/1006058.1006071}
{doi: \textsf{%
10\hspace{.1pt}\discretionary{.}{%
}{.}\hspace{.4pt}1145\discretionary{/}{%
}{/}1006058\hspace{.1pt}\discretionary{.}{%
}{.}\hspace{.4pt}1006071}}


\bibitem{westermann_real-time_1999}
R.~Westermann, L.~Kobbelt, and T.~Ertl.
\newblock Real-time exploration of regular volume data by adaptive reconstruction of isosurfaces.
\newblock {\em The Visual Computer}, 15(2):100--111, Apr. 1999. \href{https://doi.org/10.1007/s003710050165}
{doi: \textsf{%
10\hspace{.1pt}\discretionary{.}{%
}{.}\hspace{.4pt}1007\discretionary{/}{%
}{/}s003710050165}}


\bibitem{zhang_haptic_2023}
H.~Zhang, L.~Zhu, Y.~Xiang, J.~Zheng, and A.~Song.
\newblock Haptic {Rendering} of {Neural} {Radiance} {Fields}.
\newblock In {\em Proceedings of the 36th {Annual} {ACM} {Symposium} on {User} {Interface} {Software} and {Technology}}, pp. 1--10. ACM, San Francisco, Oct. 2023. \href{https://doi.org/10.1145/3586183.3606811}
{doi: \textsf{%
10\hspace{.1pt}\discretionary{.}{%
}{.}\hspace{.4pt}1145\discretionary{/}{%
}{/}3586183\hspace{.1pt}\discretionary{.}{%
}{.}\hspace{.4pt}3606811}}


\bibitem{zhou_digital_2017}
Q.~Zhou.
\newblock Digital {Elevation} {Model} and {Digital} {Surface} {Model}.
\newblock In {\em International {Encyclopedia} of {Geography}}, pp. 1--17. John Wiley \& Sons, Ltd, 2017. \href{https://doi.org/10.1002/9781118786352.wbieg0768}
{doi: \textsf{%
10\hspace{.1pt}\discretionary{.}{%
}{.}\hspace{.4pt}1002\discretionary{/}{%
}{/}9781118786352\hspace{.1pt}\discretionary{.}{%
}{.}\hspace{.4pt}wbieg0768}}


\bibitem{zhou_haptics-assisted_2008}
W.~Zhou, S.~Correia, and D.~H. Laidlaw.
\newblock Haptics-{Assisted} {3D} {Lasso} {Drawing} for {Tracts}-of-interest {Selection} in {DTI} {Visualization}.
\newblock {\em IEEE Visualization}, 2008.

\bibitem{zilles_constraint-based_1995}
C.~Zilles and J.~Salisbury.
\newblock A constraint-based god-object method for haptic display.
\newblock In {\em Proceedings 1995 {IEEE}/{RSJ} {International} {Conference} on {Intelligent} {Robots} and {Systems}. {Human} {Robot} {Interaction} and {Cooperative} {Robots}}, vol.~3, pp. 146--151. IEEE Comput. Soc. Press, Pittsburgh, 1995. \href{https://doi.org/10.1109/IROS.1995.525876}
{doi: \textsf{%
10\hspace{.1pt}\discretionary{.}{%
}{.}\hspace{.4pt}1109\discretionary{/}{%
}{/}IROS\hspace{.1pt}\discretionary{.}{%
}{.}\hspace{.4pt}1995\hspace{.1pt}\discretionary{.}{%
}{.}\hspace{.4pt}525876}}


\end{thebibliography}








\end{document}